\documentclass[aps,prb,twocolumn,superscriptaddress,citeautoscript,floatfix]{revtex4}
\pdfoutput=1
\usepackage{graphicx}
\usepackage{subfigure}
\usepackage{epstopdf}

\begin{document}

\title[Draft:\today]{Theory of vibronic assistance in the nonequilibrium  condensation of exciton polaritons 
in optically--pumped organic single crystal microcavities}

\author{Eric R. Bittner}
\affiliation{Department of Chemistry and Physics, University of Houston, Houston, TX 77204}
%\email[]{Bittner@UH.edu}

%
\author{Svitlana Zaster}
\affiliation{Department of Chemistry and Physics, University of Houston, Houston, TX 77204}

%%%\homepage[]{http://k2.chem.UH.edu}
%%%\thanks{}
%%\altaffiliation{}
%\affiliation{Department of Chemistry, University of Houston, Houston, TX 77204}
%%
%\bibliographystyle{unsrt}

\author{Carlos Silva}

\affiliation{ Department of Physics and 
Regroupement qu{\'e}b{\'e}cois sur les mat{\'e}riaux de pointe, Universit{\'e} de Montr{\'e}al,\\
C.P. 6128, Succursale centre-ville, \\
Montr{\'e}al (Qu{\'e}bec) H3C 3J7, Canada.}

\date{\today}

\begin{abstract}
We present a reaction/diffusion model for the formation of a lower polariton condensate in a micro cavity containing 
an organic semiconducting molecular crystalline film.   Our model--based upon an anthracene film sandwiched between 
two reflecting dielectric mirrors--consists of three coupled fields corresponding to 
a gas of excitons created by an external driving pulse, a reservoir of vibron states formed by the coupling between 
a ground-state vibrational model and a cavity photon, and a lower polariton condensate. We show that at finite temperature, 
 the presence of the vibron reservoir can augment the exciton population such that a lower critical pumping threshold is required to 
 achieve condensation. 
\end{abstract}

\pacs{00.00, 20.00, 42.10}
%% Keywords required only for MST, PB, PMB, PM, JOA, JOB? 
%% Uncomment for Submitted to journal title message
%\submitto{\NJP}
%% Comment out if separate title page not required
%%\pacs{}
%%\submitto{\NJP}
\maketitle
\section{Introduction}

Over the past few years, there has been considerable excitement regarding the formation of polariton Bose 
condensates in quantum dot microcavities.\cite{Kasprzak:2006mz,malpuech:21,Malpuech:2007ec,PhysRevB.67.085311,PhysRevB.72.125335,PhysRevB.81.081307,RevModPhys.82.1489,Deng:2002zt,DavidSnoke11152002,Utsunomiya:2008fk,PhysRevLett.85.2793}
This has spurred a parallel effort to observe similar effects in 
microcavity systems involving organic semiconducting films. 
\cite{PhysRevB.67.085311,chovan:045320,PhysRevB.65.195312,PhysRevLett.82.3316,somaschi:143303,lidzey:2011}
While organic semiconducting systems share 
many of the features as their inorganic counterparts, they differ in that excitons tend to be more local to the 
individual molecular sites, phonons play a central role, and the dielectric constant is considerably smaller. 
As a result, organic systems present both a challenge and opportunity for observing optical/electronic processes.
\cite{BittnerJCP2012b,C2CP23204A}

In this paper we consider the formation of polariton condensates under non-equilibrium conditions. While Bose 
condensation in atomic gases can be understood as an equilibrium phenomona since one can impose the 
condition that the average number of atoms in the gas is conserved, polaritons form due to the strong coupling 
between a photon field and an optical transition. They are transient quasi-particles that can exist only so long as 
the system is continuously resupplied with photons. Bose condensation occurs once the density of excitons 
reaches a critical threshold such that the condensate density grows exponentially in time. 

\begin{figure}[b]
\includegraphics[width=\columnwidth]{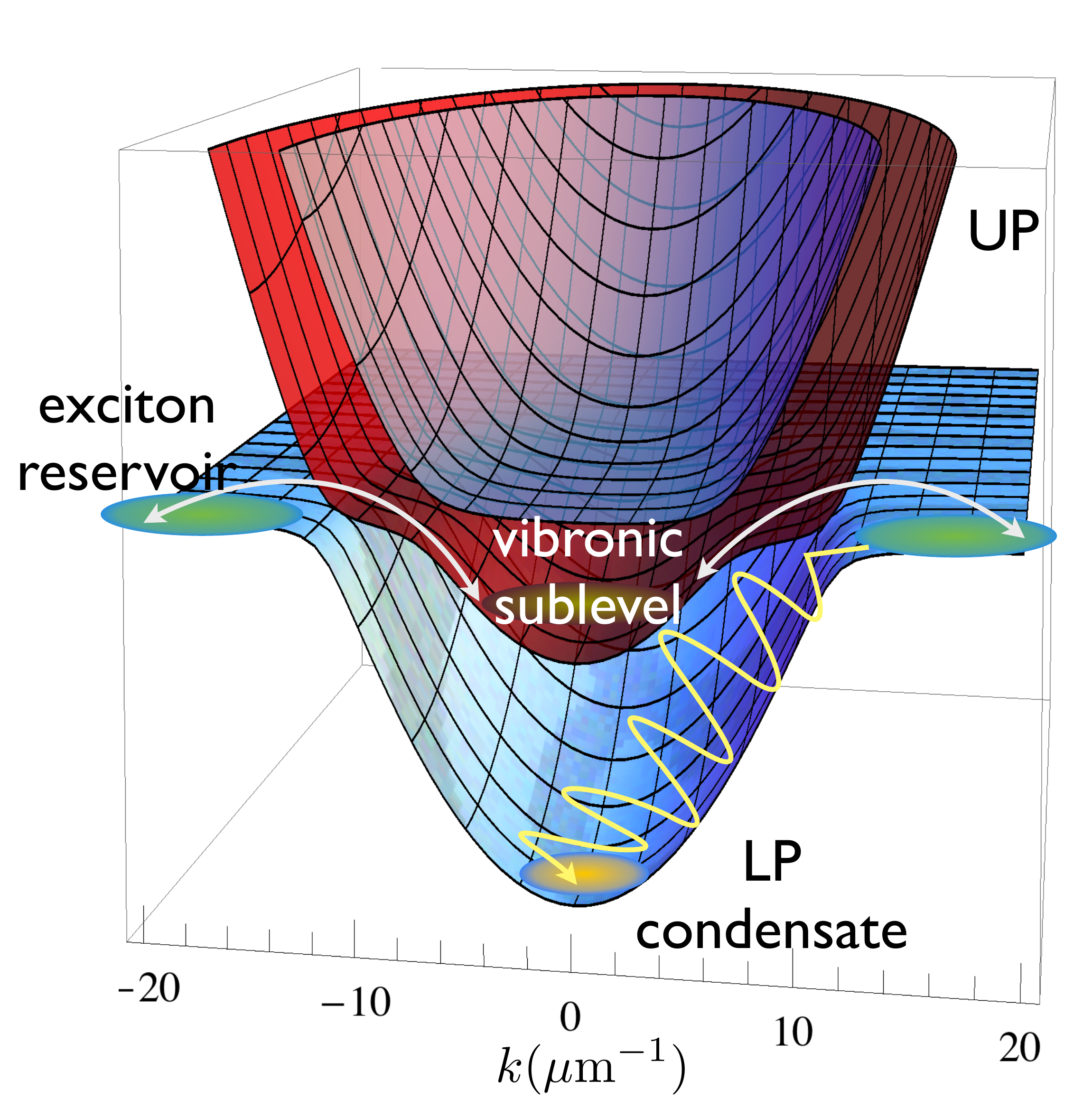}
\caption{Dispersion curves for a system with an exciton (with frequency $\omega _x$) 
and vibrational mode (with frequency $\omega _v$) is coupled to a 2D photon cavity. 
The curves are labels to reflect the dispersion of the lower polariton (LP) and two upper polariton $\left({\rm UP}
_1\right.$ and UP$_2$) branches. Here the cavity off-set, 
$\Delta $, is such that $\omega _x= \Delta  + \omega _v$ and the dashed line indficates the bare exciton 
energy. }\label{fig2}
\end{figure}

For the cases in consideration here, we assume that excitons and photons within the cavity couple to give rise to 
at least two incoherently coupled populations: free excitons and lower polaritons. 
We can also add to this picture quantized vibrational modes of the molecular electronic ground. 
%In Fig.~\ref{fig1}a, 
%we show potential curves representing the electronic ground state and 
%singlet excited state of a molecular species along with their quantized vibrational energy energy levels. 
Let us denote the vibronic states as $|n_v\rangle$ where $n$ denotes electronic excitation
 and $v$ denotes vibrational excitations.  In the weak 
coupling limit, optical transitions occur predominantly from the lowest vibrational state of the electronic ground 
state to various vibrational states in the excited electronic state, with probability weighted by the Franck-Condon 
factors. However, in the limit of strong coupling to the photon field, which is appropriate for discussing polaritons, 
the lower electronic state is dressed by the photon field and vibrational states in the electronic ground comes into 
resonance with the states in the upper electronic state. Here we consider a three-state system in which the 
ground state, $|0_o\rangle$, and a single quantized mode of the ground state, \(\left.\left|0_1\right.\right\rangle\), 
are dressed by the photon field and are strongly coupled to the \(\left.\left|1_0\right.\right\rangle\) state in the 
upper electronic state. Thus, the dressed states include the photon field in the sense that 
\begin{eqnarray}
\left\langle 0_o,n_k+1| H |0_o,n_k+1\right\rangle  &=& \hbar \omega _c(k)
\\
\left\langle 0_1,n_k+1|H |0_1,n_k+1\right\rangle  &=&\hbar  \left(\omega _v+ \omega _c(k)\right)
\\
\left\langle 1_0,n_k|H |1_0,n_k\right\rangle  &=& \hbar \omega _x,
\end{eqnarray}
where \(\omega _c(k) \approx  \left.k^2\right/2m + \Delta\) is the dispersion relation for a microcavity with cut-off 
frequency $\Delta = 2\pi c/L\eta$ and effective mass $m = 2 \pi \eta/cL$.  Here, $\eta$ is the refractive index of 
the material in the cavity and $L$ is the spacing between reflective mirrors.  For example, a $L=$2300\AA\ cavity 
containing anthracene ($\eta = 1.70$), one obtains $\hbar\Delta = 3.11$eV and an 
effective mass of $m = 1.8\times 10^{-5}m_{e}$.   A cavity with this geometry would be in resonance with the 
$S_{o}\to S_{1}$ electronic transition of anthracene.
This amounts to a ``shift'' in the lower energy  levels by $\Delta$.
%as indicated in Fig.~\ref{fig1}b.  
In other words, the presence of a photon in the cavity can bring 
any of the vibrational states into resonance with the exciton when the cavity
 itself is not in resonance with the 
 excitons.  

The coupling between the exciton state and the cavity states is given by the Rabi frequencies
$$ \hbar \Omega_1= \left\langle 1_0, n_k|H|0_0,n_k+1\right\rangle$$ and 
$$\hbar\Omega_2= \left\langle 1_0, n_k|H|0_1,n_k+1\right\rangle$$
as given by the transition moment between the $|1_{0}\rangle$ exciton state and the
two vibrational states in the ground electronic state,  $|0_{0}\rangle$  and $|0_{1}\rangle$.  
In general, the Rabi frequency for an on-resonance excitation can be expressed 
in terms of the electric field amplitude of the cavity mode and the transition dipole moment viz.
\begin{eqnarray}
\Omega = \mu_{fi}E/\hbar
\end{eqnarray}
If we invoke the Franck-Condon principle and assume that the transition occurs in a fixed 
frame of the nuclear motion, then 
\begin{eqnarray}
\mu_{fi} = \mu_{10}f_{0-\nu}
\end{eqnarray}
where $f_{0-\nu} \le 1 $ is the overlap integral between displaced harmonic oscillator states and $\mu_{10}$ is the 
exciton transition moment.  Thus, knowing the Rabi frequency, $\Omega_{1}$ for the exciton in the cavity we can reliably estimate 
\begin{eqnarray}
\Omega_{2} = f_{0-1}\Omega_{1} = f_{0-1}(f_{0-0}\mu_{10}E/\hbar). \label{hrfact}
\end{eqnarray}
The overlap is determined by the Huang-Rhys factor, $S$ with $f_{0-v} = \sqrt{S^{v}/v!} \exp(-S)$.  For polyacene 
systems such as anthracene, the Huang-Rhys factor for the high-frequency C=C modes between the $v = 0$ vibrational state of the exciton 
and the $v = 1$ vibrational state of the ground electronic state is approximately $S = 1$ for the lowest excited state
 as determined by comparing the (0-0) and (0-1) spectral areas.\cite{yamagata:204703,JR9600005206}  This implies that
 $\Omega_{2} \approx 0.36\times\Omega_{1}$.   
  
Diagonalizing  $H$  in the dressed basis, one obtains polariton dispersions similar to what is shown in Fig.~\ref{fig2}. 
Here, we have taken the exciton energy at 3.11 eV, which is the $S_o\rightarrow S_1$
transition in anthracene, included a single vibrational mode at 0.2 eV, and set the cavity cut-off at 2.9 eV so that 
$\Delta = \omega _x-\omega _v$.  
  The resulting polariton states are superpositions of a cavity photon mode 
with the molecular exciton and ground-state vibrational mode (which we'll refer to as a ``vibron'' mode).
The lower polariton (LP) branch, which is predominantly ``photon'' in character undergoes  avoided crossings with the two
upper polariton branches which are predominantly vibronic or excitonic in character.   

A  multi-state model involving the mixing between the cavity and the vibronic sub levels was used recently by 
K{\'e}na-Cohen and Forrest\cite{kena-cohen:073205,Kena-CohenS.:2010fk} to describe the multiple resonances observed in 
anthracene microcavites.  In this case, the vibronic sublevels are those of the electronic excited state and not 
the ground state, which we assume.   At thermal equilibrium, all of the population 
is in the  $|0_{0}\rangle$ state and the vibrationally excited states would have essentially no population and would 
not ordinarily come into play in the optical response of the system.  However,  in a driven system and one in which 
the $|1_{0}\rangle$ exciton state does carry oscillator strength to the $|0_{\nu}\rangle$ states 
via the Franck-Condon principle, one should be able to create a non-thermal {\em population inversion } in these states
which can act as a reservoir.

 In this paper, we consider the role that vibrational cooling may have on the formation of a polariton condensate in 
the \textit{ incoherent} pumping regime. In a recent set of experiments by Lidzey{'}s group, it was apparent that 
vibrational modes can contribute to the LP polariton population and may serve as an important cooling mechanism for UP and free excitons. 
\cite{lidzey:2011,somaschi:143303}  
Our model consists of a reaction/diffusion equation whereby a population of free excitons is coupled to a population of 
vibrational modes and can serve as a reservoir for the formation of a LP Bose condensate.  Our model assumes that 
condensation occurs when fluctuations about a steady state condensate with zero initial amplitude begin to grow exponentially 
in time once a threshold population of excitons has been achieved.  The model results in a series of phase-diagram
of exciton pumping rate vs. lattice temperature and cavity cut-off.  We show that when the cavity cut-off resonance from the 
exciton mode, thermalized vibrons can enhance the exciton population and lead to lower critical pumping rates as the lattice 
temperature is increased.  This seems counterintuitive since one expects  condensates to form as the temperature is lowered. 
However, as we discuss next, the symmetry breaking transition in a driven system comes about due to the 
external driving force rather than through internal interactions or density of states. 

\section{Theoretical reaction/diffusion model}
We begin with the requirement that the condensate density obey a continuity equation with sink and source terms 
representing the temporal decay of the condensate with effective mass $m_{LP}$ due to cavity loss (with rate $
\gamma $) and creation of new condensate due to the presence of a singlet exciton reservior, $S$. Thus, 
we write using the hydrodynamic form \cite{Holland:1993}
\begin{eqnarray}
\partial _t\rho  = - \nabla \cdot (\rho \nabla s)/m_{LP}+ (r\cdot u - \gamma )\rho ,
\end{eqnarray}
where the action, $s$  satisfies a quantum Hamilton-Jacobi equation of the form:
\begin{eqnarray}
\partial _ts + \frac{1}{2 m}(\nabla s)^2 - \frac{\hbar ^2}{2m}\frac{1}{\sqrt{\rho }}\nabla ^2\sqrt{\rho } + g \rho  = 0,
\end{eqnarray}
and that the current $ J  = \rho \nabla s/m$. 
Using $\phi= \sqrt{\rho }\exp(i s/\hbar)$ as the condensate amplitude, one obtains a Gross-Pitaevskii (GP) 
equation that properly incorporates the source and sink terms as an optical potential term. 
\begin{eqnarray}
i \partial _t\phi = -\eta\nabla ^2\phi + \left(g|\phi |^2+\frac{i}{2}(r \cdot S-\gamma  )\right) \phi
\label{gp}
\end{eqnarray}
where $\eta = {\hbar}/{2m_{LP}}$ is the LP diffusion constant and $g$, $r$, and $\gamma$ are rates.
If we take the exciton reservoir to be constant (in both time and space), then we can begin to see the conditions 
necessary for establishing a condensate in a driven, non-equilbrium system. 
The trivial solution is given by $\phi = 0$  in which no condensate will forms at any time. 
This provides us with the trivial steady-state solution of  $\phi _{ss}(x,t) = 0$.

However, if we introduce non-trivial fluctuations about the steady state, 
$\phi(x,t) =\phi_{ss}(x,t) + \delta \phi(x,t)$  and ask whether or not such amplitude fluctuations will grow or decay 
in time,  one sees that if the optical potential term $(r \cdot S-\gamma ) <0 $, 
then any fluctuation in the polariton component will decay exponentially
in time and no long-lived condensate will form. 
On the other hand, if  the exciton density is sufficiently large that $(r\cdot S - \gamma ) > 0 $, 
then the condensate amplitude will grow exponentially with a rate $(r \cdot S -\gamma )$. 
Thus we conclude there exists a critical exciton density, $S_o$  whereby the polariton decay due to cavity leakage is 
exactly counterbalanced by repopulation from a free exciton reservoir. 
Once a critical density of excitons has been achieved in a given region, 
the LP condensate population  in that region experiences exponential growth.  Thus, one expects that will occur once
$S_{o} = \gamma/r$.  Taking the exciton density to be proportional to the exciton pumping rate, $p_{o}$, then one expects
$
p_{o,crit} \propto {1}/{r}.
$
The presence of a threshold exciton 
population for the formation of a condensate at long time is the hallmark of a quantum phase transition in a 
non-equilibrium system. \cite{PhysRevLett.96.230602,PhysRevLett.107.080402,PhysRevLett.99.140402}
The principle result of the work presented here is that the inclusion of dressed ground state vibrational modes

%\section{Polariton condensation: role of vibronic cooling}

As discussed in the Introduction, local vibrational modes play very important roles in the electronic spectroscopy of 
most molecular semiconductors and  in  microcavity system, may serve as a viable cooling mechanism for excitons.
With this in mind, we adopt a continuum model where we 
write the singlet exciton density, $S(x,t)$, and vibron density, $v(x,t)$, as a solutions to the following 
reaction/diffusion equations which, in turn, are coupled to the GP equation for the condensate amplitude,
\begin{eqnarray}
\partial_{t}S &=& D_{s}\nabla^{2}S - r\cdot S | \phi(x,t)|^{2} + p(x,t) \nonumber \\ &-& \gamma_{S}S + (k_{1} v(x,t) - k S) \\
\label{ex}
\partial_{t}v &=& D_{v}\nabla^{2}v - k_{1} v(x,t)  + k S\label{vib}
\end{eqnarray}
where $r S|\phi|^2$ corresponds loss of exciton density in a given spatial region due to the formation of the condensate, $\gamma_{S}$
is the singlet exciton radiative rate, and $p(x,t)$ is the rate that excitons are pumped into the system by some external source. 
We imagine a experimental situation where excitons in a thin-film within a microcavity
are created in a given region illuminated by a steady laser spot, the excitons diffuse incoherently through
through the sample as UP polaritons and are either rapidly thermalized with UP vibrons or decay through radiative process. 
Taking ${\cal A} = \pi\sigma$ as the area illuminated by the pumping pulse we can relate the pumping rate to an irradiance
via 
 $E_{e} = p_{o}\omega_{x}/{\cal A} \approx 0.32  p_{o} {\rm W}/{\rm m}^{2}$ 
taking $\sigma = 500 \mu{\rm m}^{2}$ and $p_{o}$ is given in ${\rm ns}^{-1}$. 
Thus a $p_{o} = 4000{\rm ns}^{-1}$ would correspond to an irradiance of 1.3 ${\rm kW}/{\rm m}^{2}$ which is
comparable to the average solar irradiance on the Earth's surface. 

Putting aside any decay channels for the moment, let us assume that the
exciton and vibron populations are rapidly thermalized so that we can write
$$k_{1} = e^{-\beta \delta} k$$
where $\delta = \omega_{x}-(\omega_{v} + \Delta)$ is the energy gap between the 
exciton and vibrons.  This provides a useful means for incorporating the 
geometry of the cavity and the temperature of the lattice itself into our equations of motion.  
We also have assumed that the energy difference between the free exciton reservoir and LP 
polariton condensate is much greater than $k_{B}T$ so that any exciton population removed by condensate formation does not escape to repopulate the free exciton reservoir.  A final assumption we make is that the diffusion constant appearing in Eq.~\ref{vib} can be effectively set to zero since we can assume that  molecular vibrational excitations are local to the molecular sites in the system.  

% Lacking firm experimental values for the various rates, we assume that these are  $\approx 10  \gamma_{S}$.
%Fluctuations in the condensate density introduce a bi-molecular channel for the loss of excitons in a given spatial region, which can be 
%replenished through exciton diffusion back into that region, thermalization of the vibron reservoir, and by further pumping by the external field.  

\begin{figure*}[t]
\subfigure[]{\includegraphics[width=0.66\columnwidth]{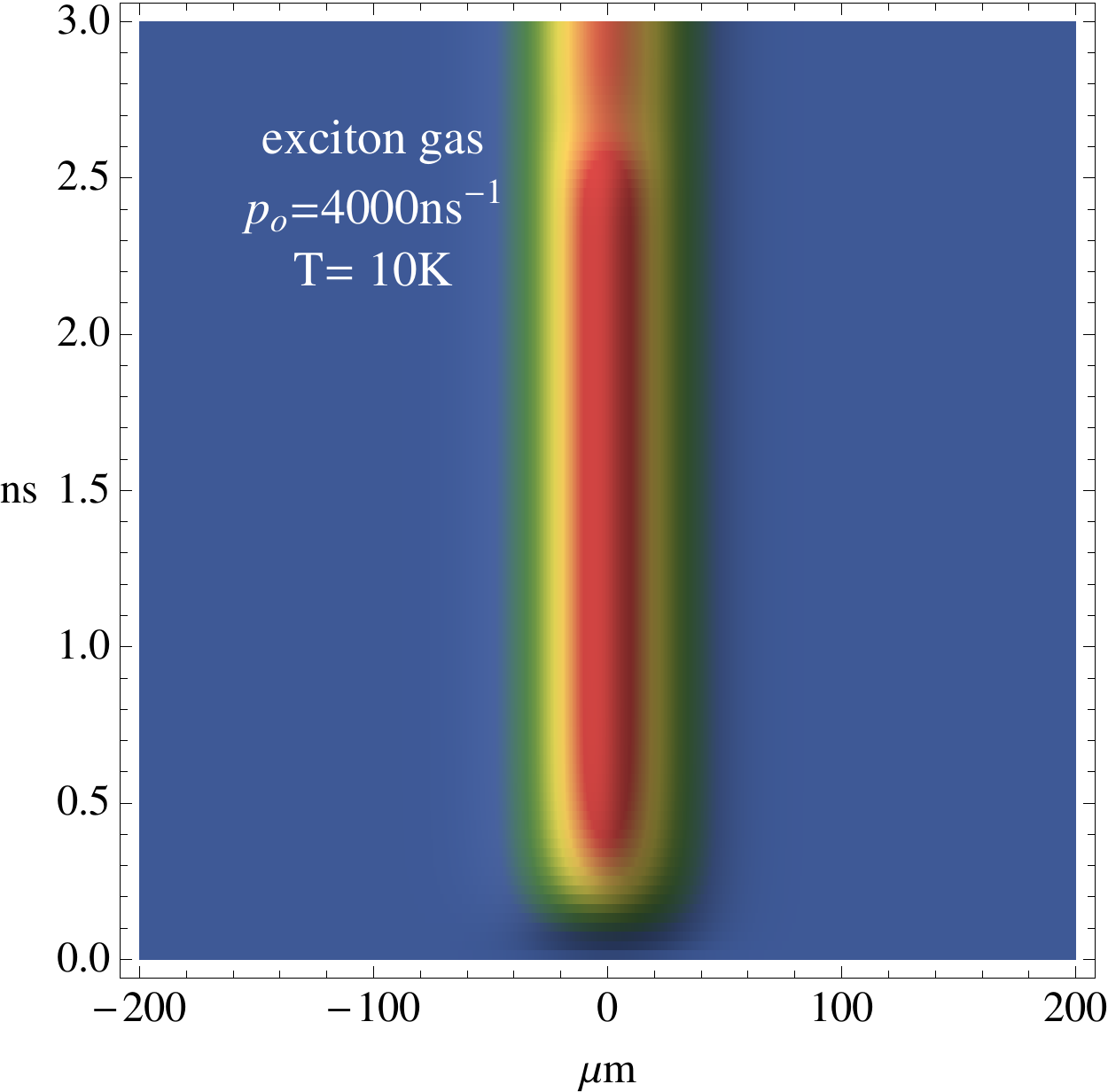}}
\subfigure[]{\includegraphics[width=0.66\columnwidth]{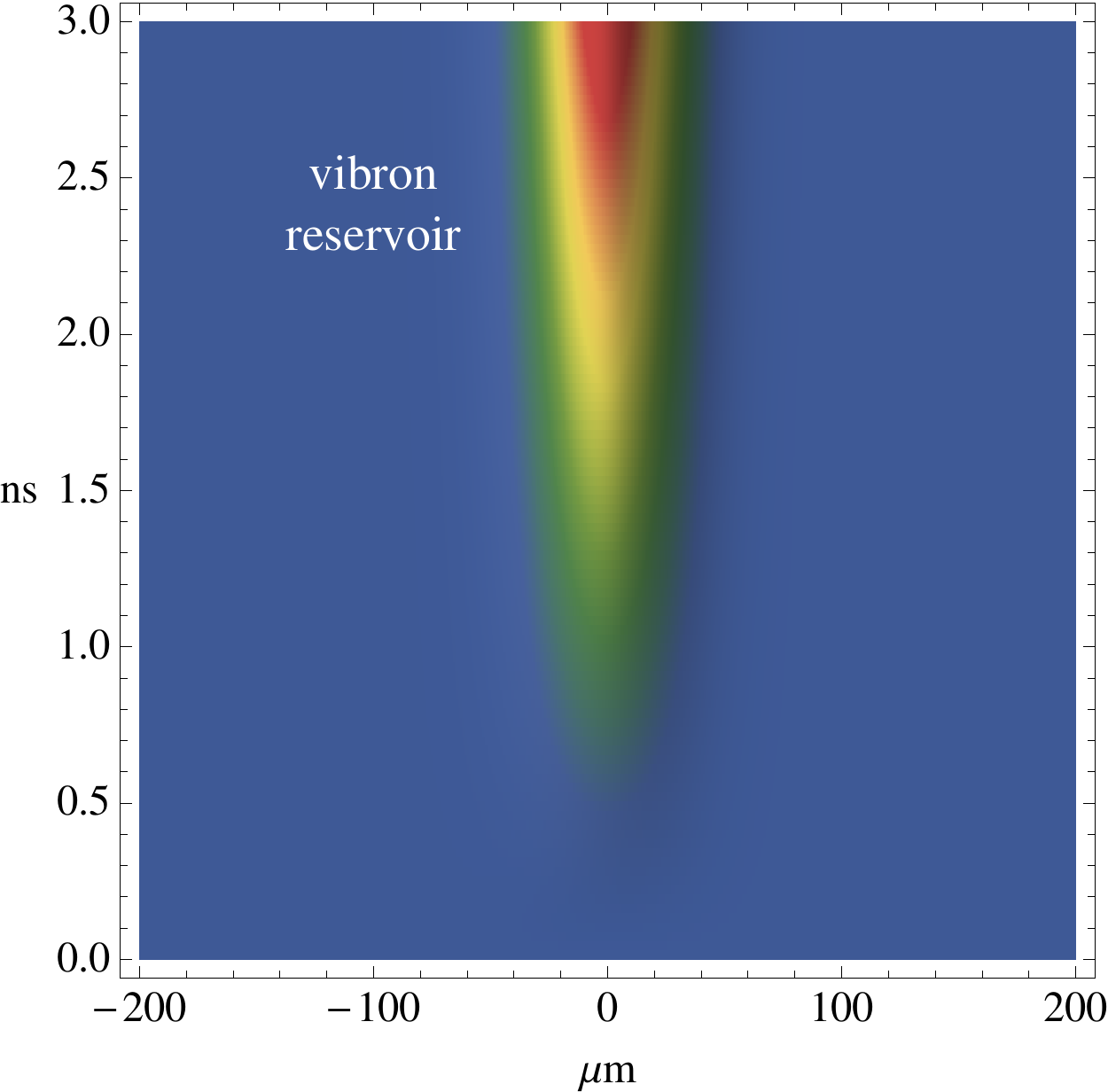}}
\subfigure[]{\includegraphics[width=0.66\columnwidth]{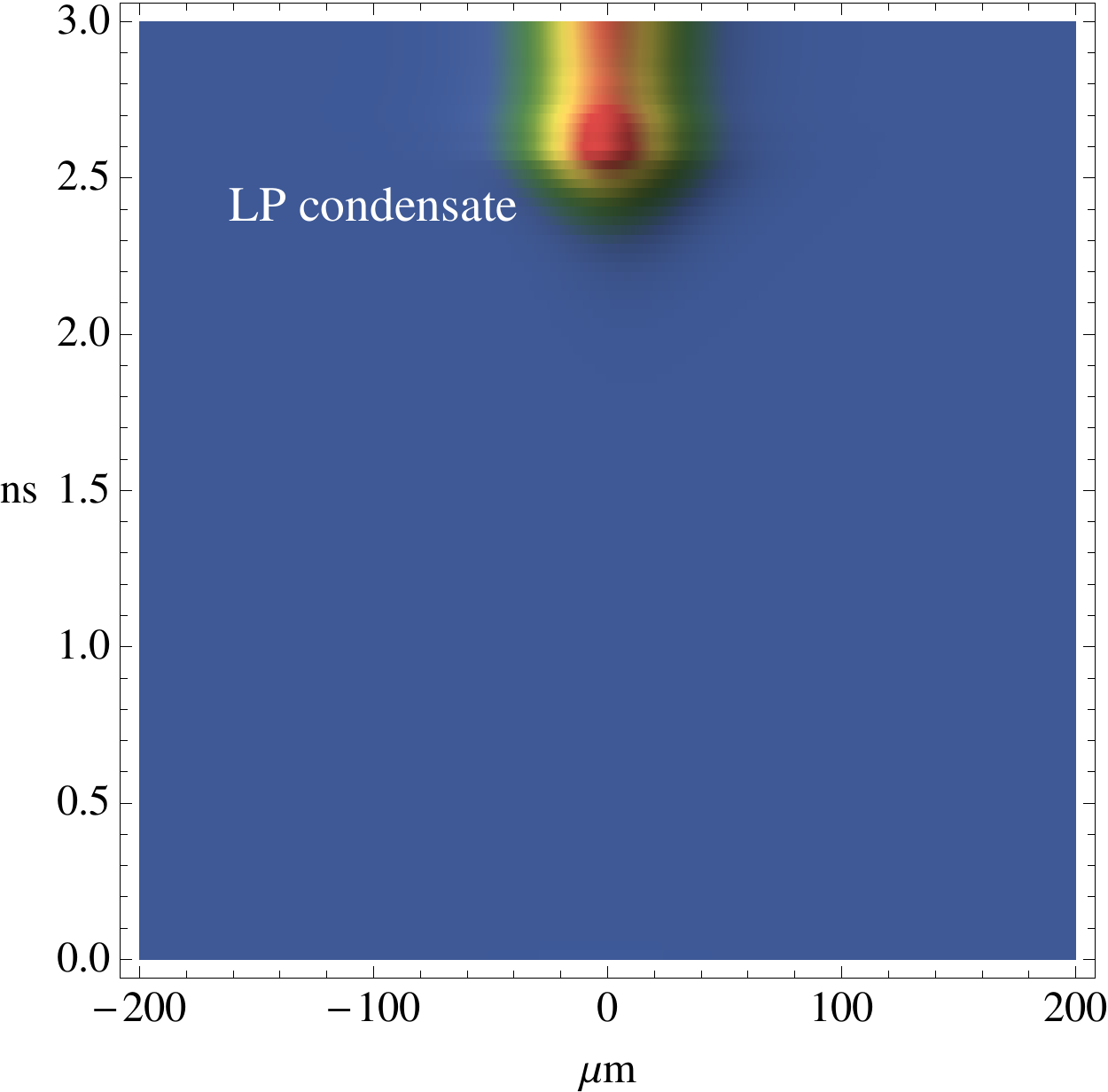}}
\subfigure[]{\includegraphics[width=0.66\columnwidth]{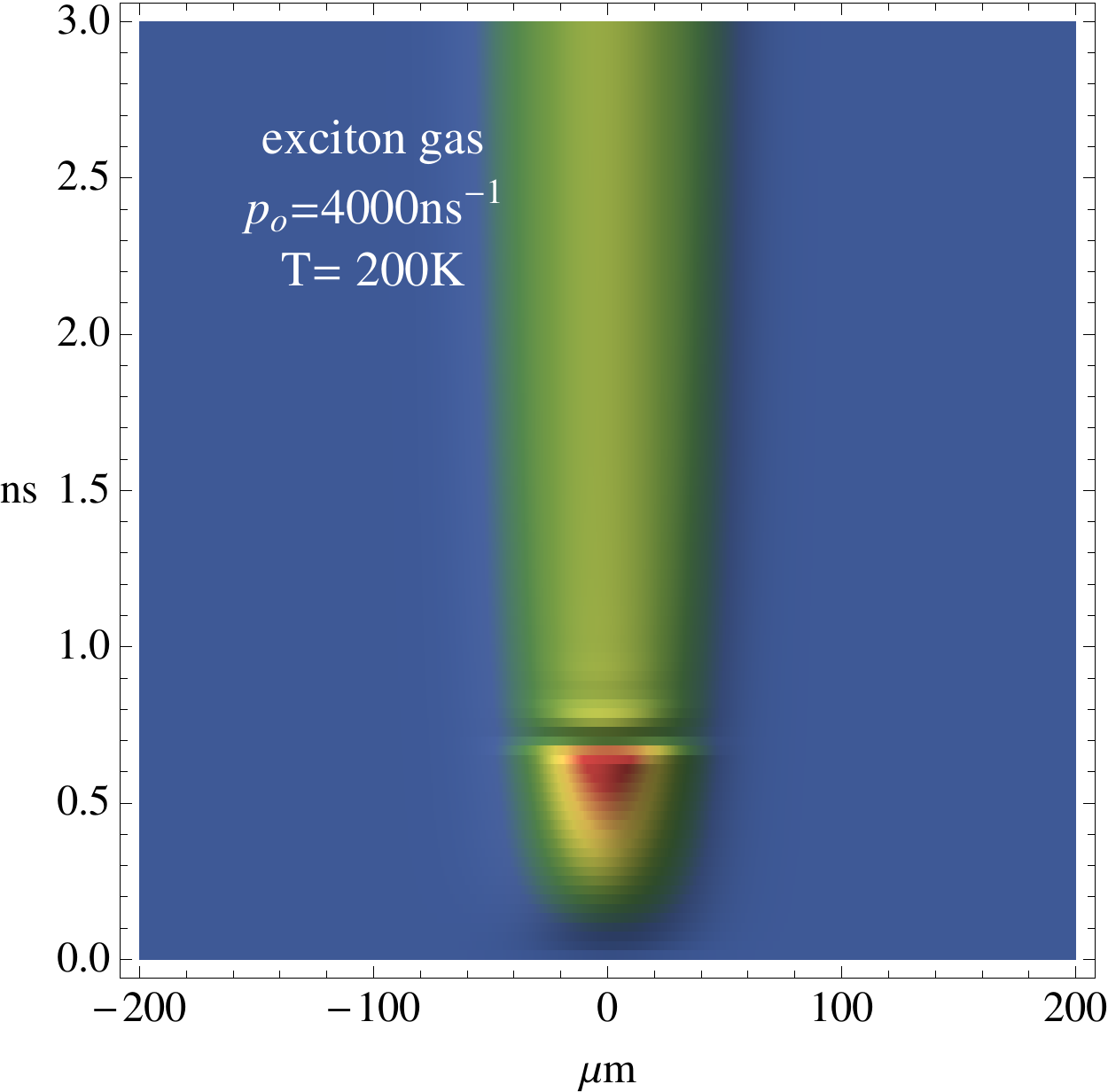}}
\subfigure[]{\includegraphics[width=0.66\columnwidth]{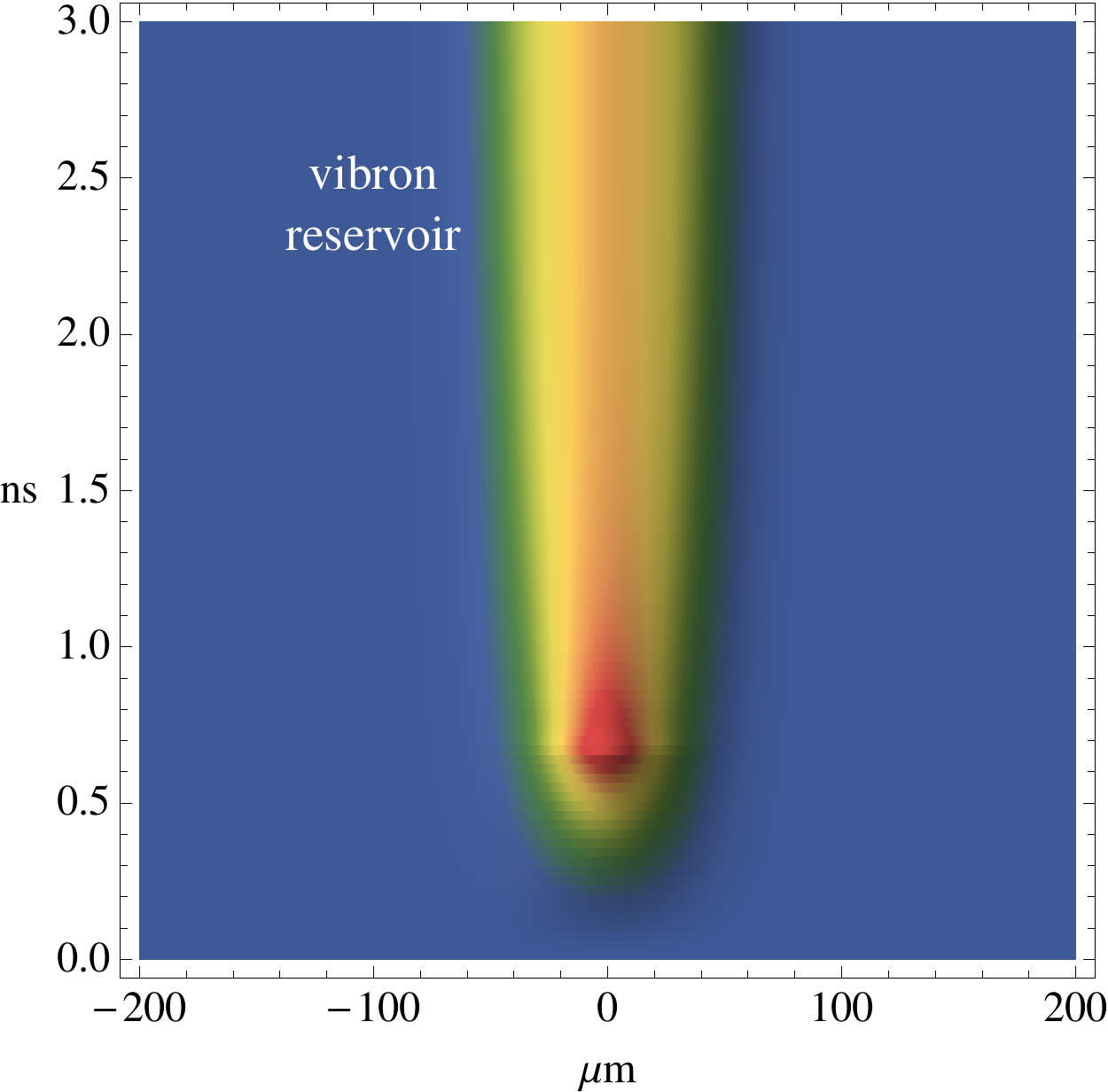}}
\subfigure[]{\includegraphics[width=0.66\columnwidth]{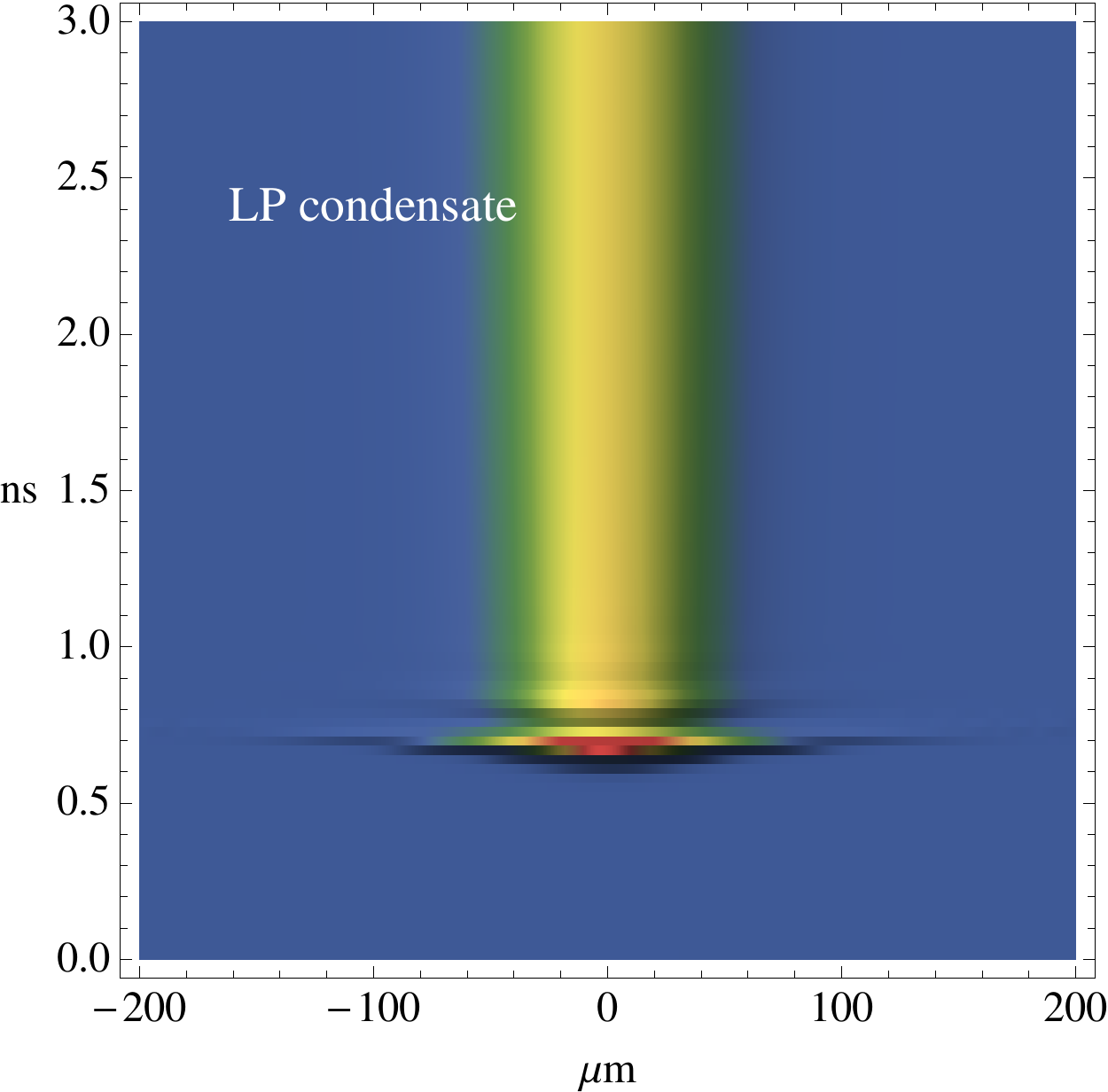}}
\caption{\label{fig:solns} Exciton gas, vibron reservoir, and polariton condensate formation at $p_o = 4000{\rm ns}^{-1}$
in  detuned cavity ($\Delta = 2.88{\rm eV}$, $\omega_{x} = 3.1{\rm eV}$, and $\omega_{v} = 0.2{\rm eV}$). 
 (a-c) $T = 10 K$ (d-e) $T = 200K$.  }
\end{figure*}

%One of the shortcomings of our model is that we do not have a precise handle on various rates used as input. 
%In the model calculations presented in this paper we  
%used the two exciton decay rate constants $\gamma_{S} = k$,
%to set the time-scale and used the singlet exciton diffusion constant 
%to set the length-scale of our model.  $\ell = \sqrt{D_{S}/\gamma_{S}}$.
%Taking $\gamma_{S} = 0.1 {\rm ns}^{-1}$ and  the experimental 
%singlet exciton diffusion constant  $D_{S} = 0.13 \mu{\rm m}^{2}/{\rm ns}$ at 10K\cite{PhysRevLett.41.131,PhysRevB.31.2430}, 
%$ \ell = 1.14 \mu{\rm m}$. 
% Taking the area of the pumping pulse to be $A \propto \pi\sigma = 100\pi \ell^{2}  = 408 \mu{\rm m}^{2}$ and
% $p_{o} \approx 0.2 \times \gamma_S$ 
%an exciton energy of $\omega_{x} = 3.1$ eV, and $p_{o} \approx 0.2 \times \gamma_S$ (from Fig.~\ref{fig3}), an
%irradiance (in the cavity) on the order of  $E_{e} \approx 1$W/m$^{2}$ 
% would be required to achieve the free exciton density needed to produce an LP condensate.  
% We also assumed that the polariton decay rate constant, $\gamma$ in Eq. ~\ref{gp} is small compared to the $\gamma_{S}$.
%In the results presented in Figs.~\ref{fig3} and \ref{fig4}, we take $\gamma = 0.01 \gamma_{S}$.  
%Assuming the UCT and LCT rates increase linearly with $\gamma$,  

\section{Numerical results}

%The details of our model and its parameterization is given in the appendix (or supplementary material) of this paper.  In brief, we 
%attempted to use as much experimental data as possible in setting the diffusion constants and rates that appear in our equations of motion. 
%All numerical simulations were performed using the {\em method of lines} approach as implemented in Mathematica.\cite{MMa8.0}
%
%

We next consider the numerical integration of the equations of motion given above.  For this, we use the method of lines approach 
as implemented in the NDSolve[] routine in Mathematica.\cite{MMa8.0}
For numerical purposes, we use the exciton decay rate, $\gamma_{S}$ and the singlet exciton diffusion constant to set the time and 
length-scales of our model. The exciton decay rate, $\gamma_{S}$ includes all radiative and non-radiative processes that diminish the exciton population that do not contribute to the formation of either lower polaritons or $|0_{1},n+1\rangle$ vibrons.  Since the singlet exciton radiative lifetime is at least 1 ns,  
setting $\gamma_{S} = 1{\rm ns}^{-1}$ is a reasonable estimate. 
 From transient singlet diffusion constant for excitons in anthracene is 
$D_{S} = 0.13 \mu{\rm m}^{2}/{\rm ns}$ at 10K\cite{PhysRevLett.41.131,PhysRevB.31.2430}.  This produces a length-scale 
$\ell = \sqrt{D_{S}/\gamma_{S}} = 1.14 \mu{\rm m}$.  From Eq.~\ref{pump}, excitons are pumped into the system  in an area
 ${\cal A} \propto \pi\sigma = 100 \pi\ell^{2}$, so that $p_{o}/{\cal A}$ gives the number of excitons that are created in the system per unit time per unit area.  
 
 Below, we consider two possible regimes.  One in which lifetime of the exciton gas is short relative to that 
 of the polariton condensate and the other in which the exciton gas is long-lived.  In each case, we ``seed'' the condensate 
 with a small amplitude about the center of our grid and determine whether or not this ``seed'' fluctuation grows or decays
 as the temperature and pumping strength are varied.

\subsection{Slow interconversion regime}

Let us first consider the case where the conversion from the exciton gas to the LP condensate is slow compared to the exciton to 
vibron conversion.  We shall refer to this as a slow conversion regime and take $r << k$ in our equations of motion.  
We would expect this 
limit to be valid when the energy difference between the vibron state and the free exciton 
is on the order of $kT$ and the Rabi splitting between 
the exciton and cavity modes to be small.
Taking the pumping rate to be a gaussian that grows exponentially to some maximum value
\begin{eqnarray}
p(x,t) = p_{o} e^{-x^2/2\sigma}(1-e^{-t/\tau})\label{pump}
\end{eqnarray}

In Fig.\ref{fig:solns} we show the results of a representative numerical simulation using a pumping rate of $p_{o} = 4,000 {\rm ns}^{-1}$, 
which corresponds to an irradiance of $\approx$ 3.1kW/m$^{2}$, which is above the threshold for forming 
the LP condensate. As point of reference, solar irradiance of the earth is on the order of 
1 ${\rm kW}/{\rm m}^{2}$.  The three panels in Fig.\ref{fig:solns}(a-c) are the time-dependent densities for the 
free exciton gas, the vibron reservoir, and the condensate at a lattice temperature of 10K. As seen in Fig.~\ref{fig:solns}a, the exciton gas 
forms and begins populating the vibron reservoir.  As the population in the vibron reservoir increases, it augments the 
exciton density such that a critical population can be established for forming a stable condensate.  

In Figs~\ref{fig:solns}{d-f} we show the formation of the condensate when the lattice temperature is considerably warmer at 200K.  
Here the vibron reservoir rapidly augments the exciton population and the condensate forms at a much earlier time than in the previous 
example at 10K.  Also, one notes that the condensation is far more abrupt 
and produces a rapidly evolving condensate wave packet. The system does, however, reach a steady state at long time.

\begin{figure*}
\subfigure[$\Delta = 2.85$eV]{\includegraphics[width=0.6\columnwidth]{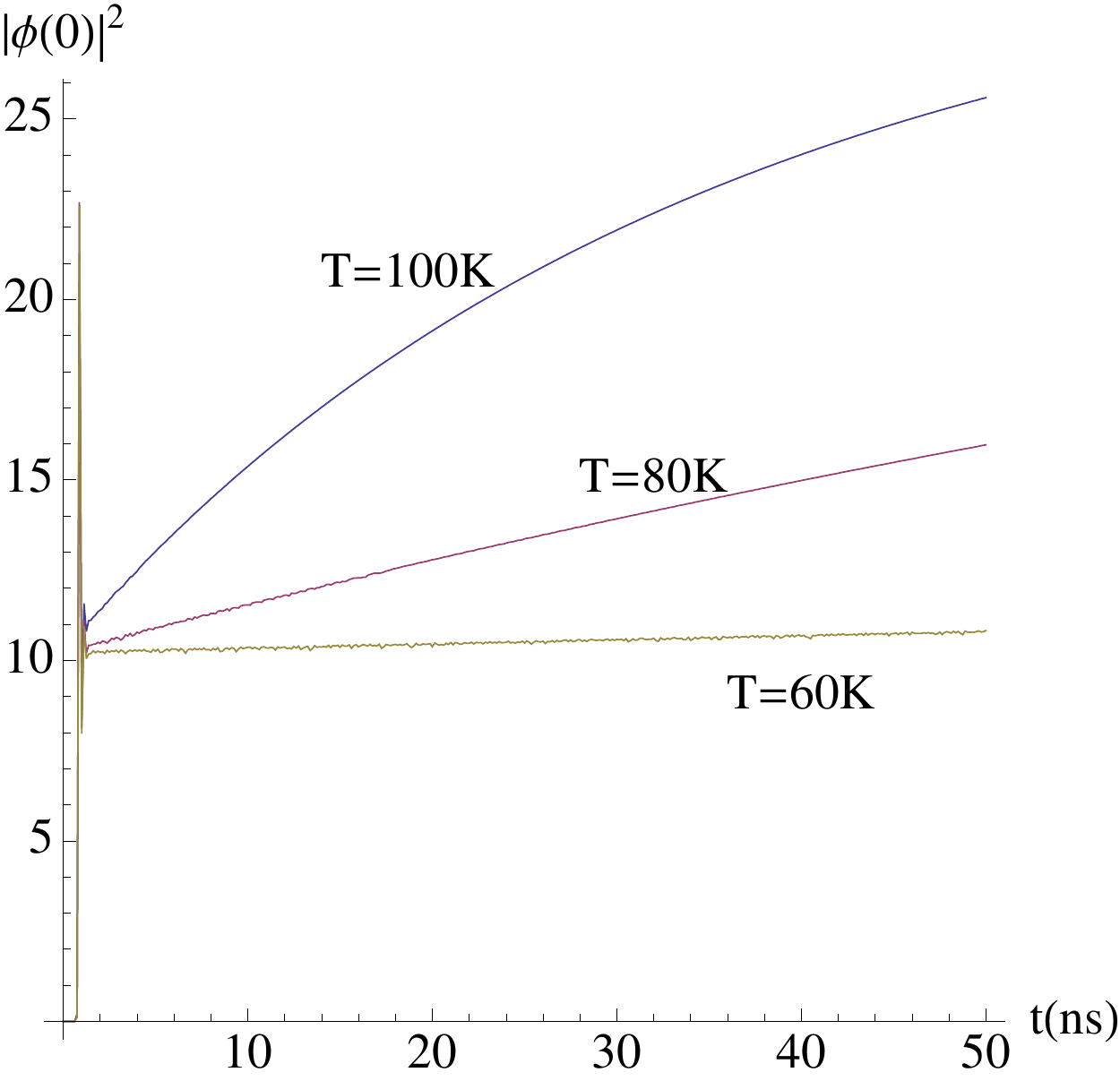}}
\subfigure[$\Delta =2.88$ eV]{\includegraphics[width=0.6\columnwidth]{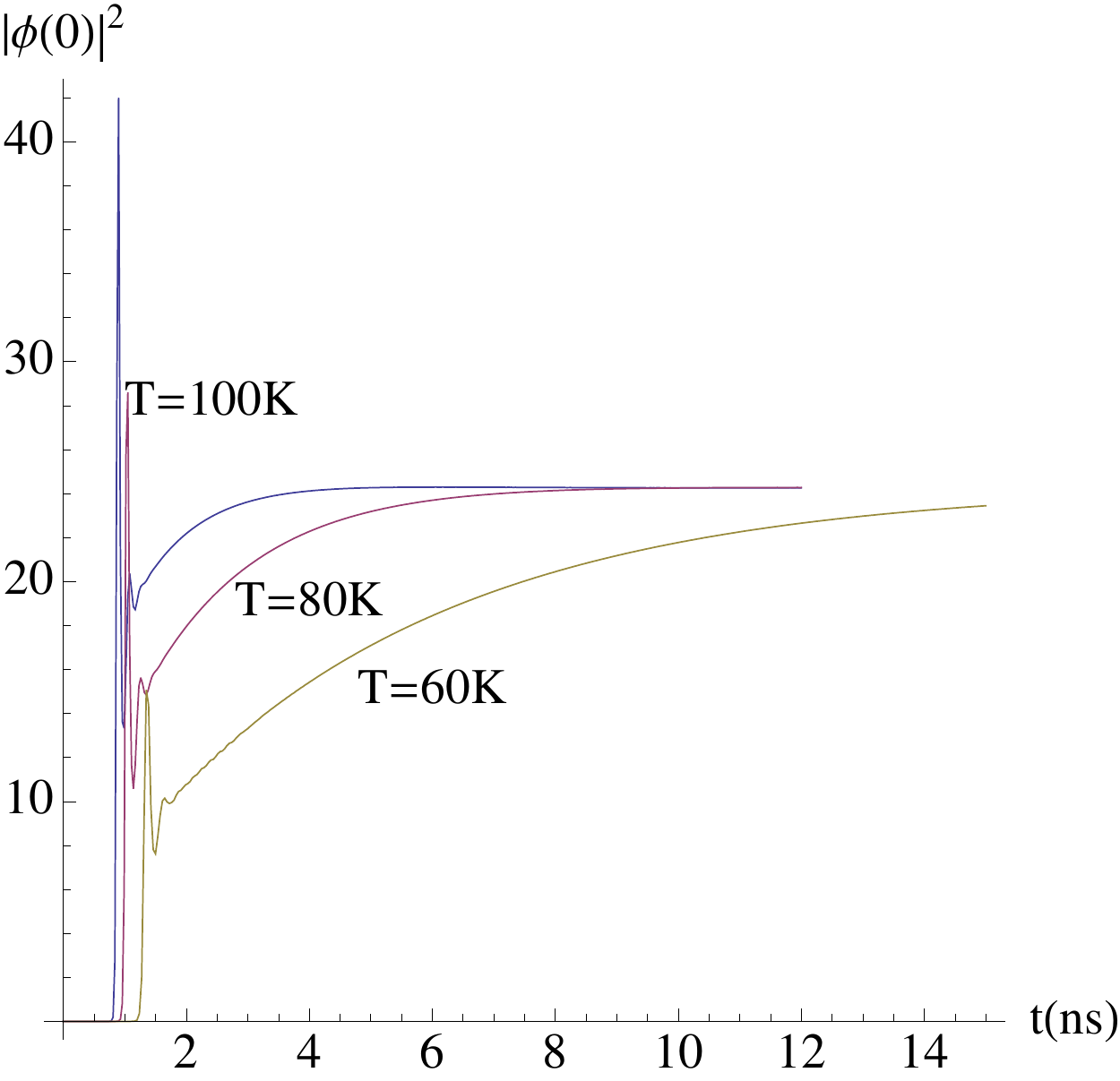}}
\subfigure[$\Delta =3.0 $eV]{\includegraphics[width=0.6\columnwidth]{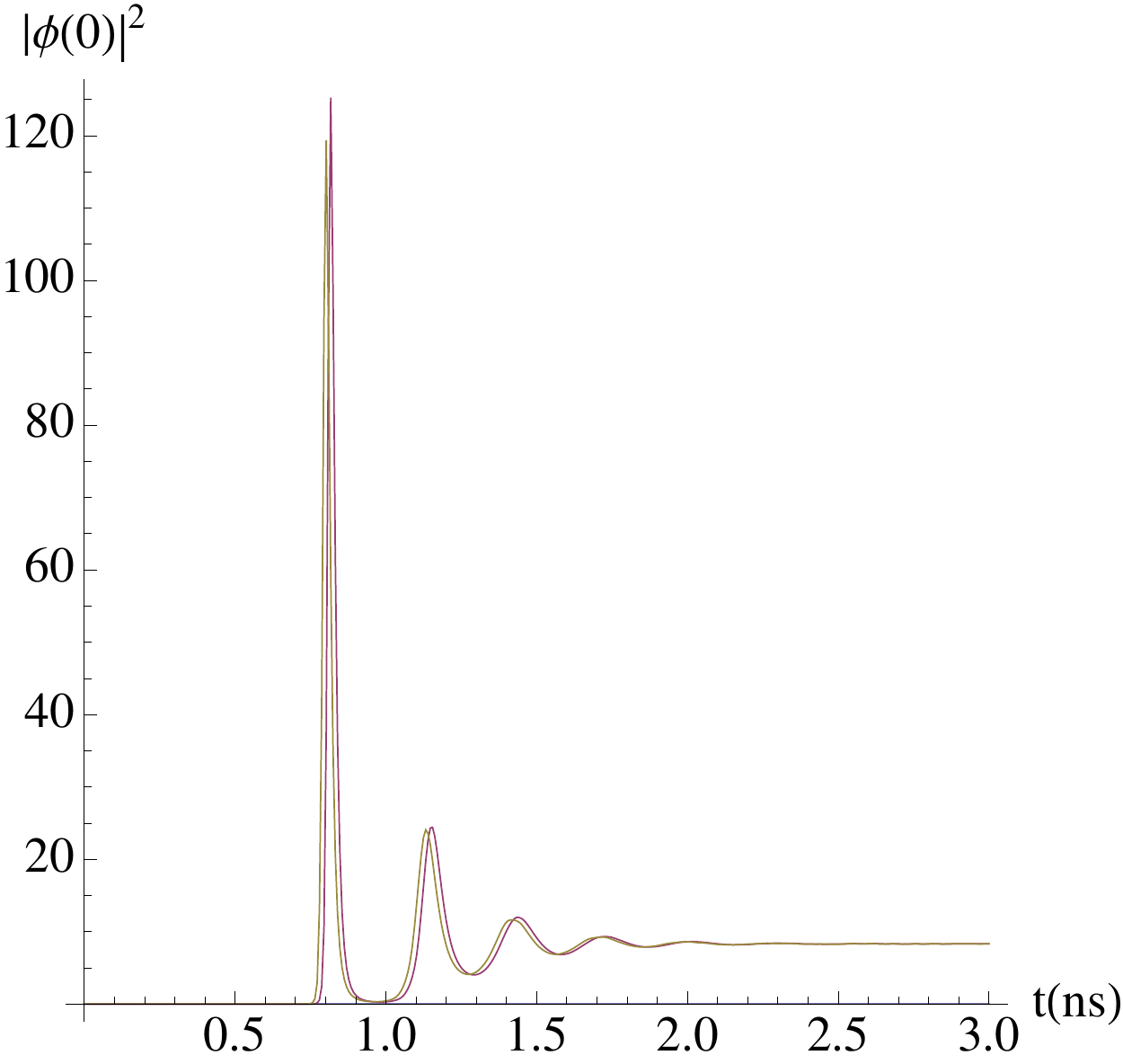}}
\caption{Condensate density $|\phi(t)|^{2}$ at $x = 0$ for different cavities for $p_{o}$ above the critical pumping threshold. }\label{fig4}
\end{figure*}

Once the condensate forms, it depletes the exciton density and at long times, the system (exciton gas, vibron reservoir, and condensate)
approaches a steady state. This is evidenced in Fig.\ref{fig4}(a-c) were we show the condensate density at $x = 0$ (center of the grid) versus time
for various cavities at different temperature.  In each case, the exciton pumping rate is above the critical threshold.   

In Fig.~\ref{fig4}a and b, the vibron 
state (at $\Delta  + \omega_{v}$)  is lower in energy than the exciton energy.   In both cases, the condensate forms about 1 ns after the pumping is initialized.  
In all cases, the initial rapid buildup of polariton density drops almost immediately due to the ballistic spread of the polariton wave packet away from the pumping region. 
After one or two ``bounces'' the polarition density relaxes to a steady state population.  

In Fig.~\ref{fig4}a, the cavity off-set at $\Delta = 2.85$eV is such that the
vibron level is considerably lower than the exciton level. In this case, the initial formation time of the polariton condensate shows a very weak dependency on the 
temperature, but the steady-state population is clearly dependent upon the temperature with warmer systems leading to a greater steady state polariton density at long times.  
In the intermediate case at $\Delta = 2.88$eV, the polariton steady state is reached much sooner. While the rate at which the polariton population approaches the
steady state does depend upon the temperature, the final steady-state population shows very little dependency.

Lastly in Fig.~\ref{fig4}c we consider a case in which the vibron level is well above the exciton level and plays no significant role in the 
formation of the polariton condensate at any temperature.  Here, we can again see a series of beats as the polariton condensate forms, decays, and reforms
until a steady-state population is achieved.  These beats are not due to quantum coherences between the exciton and polariton since we
are treating the exciton gas as a classical reservoir.  The beats are are due to population depletion and replenishment from both the 
vibron reservoir and the pumping field.  

\begin{figure}[b]
\includegraphics[width=\columnwidth]{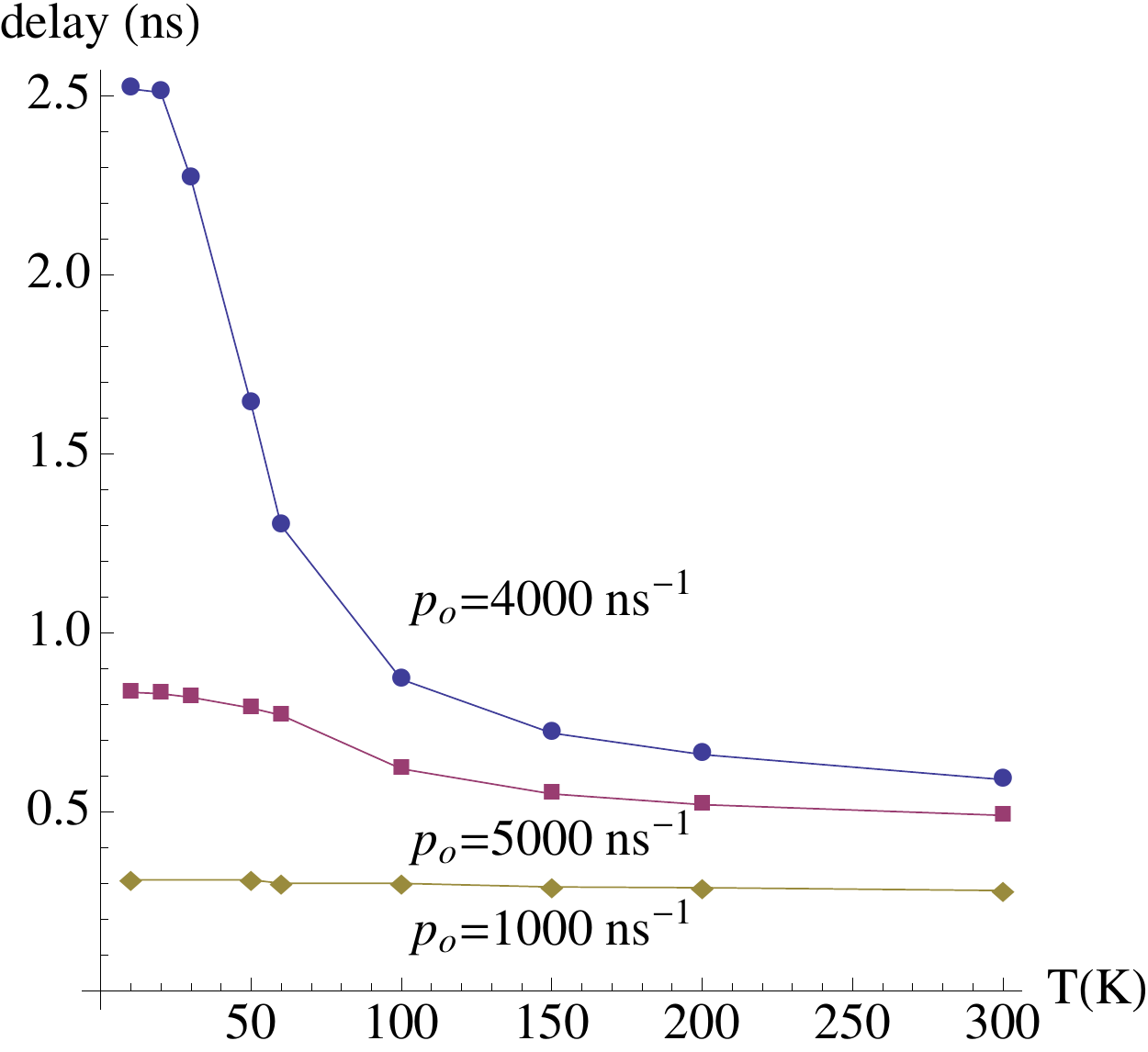}
\caption{Delay of condensate formation in a ``standard'' anthracene cavity model.
($\Delta = 2.88{\rm eV}$, $\omega_{x} = 3.1{\rm eV}$, and $\omega_{v} = 0.2{\rm eV}$).  }\label{fig3}
\end{figure}

In examining the final steady state populations, we note that steady-state condensate population is enhanced by at least a 
factor of 3 to 5 by the presence of the vibron reservoir.  This implies that cavities which can take advantage of the lower-lying vibron levels
can produce a higher density condensate which translates into a more intense steady-state optical signal.  

% correct this paragraph
In Fig.~\ref{fig3} we show how the condensation delay times  depend upon both the lattice temperature and the exciton pumping rate for a cavity with 
$\Delta = 2.88$eV.  The model does give a critical pumping rate of  $p_{o} = 3800 {\rm ns}^{-1}$, corresponding to an 
irradiance of  1.2 kW/m$^{2}$, below which long-lived polariton condensates will not form 
at any temperature.  Close to this threshold, the delay time is very sensitive to the lattice temperature to the extent that 
for a very cold lattice the condensation delay time can be $\approx 6 \times$ longer that of a warmer lattice.  It is important to point
that an irradiance of 1.2 kW/m$^{2}$ is essentially the intensity of bright sunlight which implies that the critical pumping rates 
given by our model are not unreasonably intense and can easily be achieved.  However, this is the irradiance {\em within} the cavity itself
and we have not taken into account the fact that the pumping laser needs to penetrate through the DBR to the cavity. 

On first thought it seems counterintuitive that a {\em warmer} system would have an easier time forming a stable polariton condensate since we expect condensation to occur as the temperature is lowered past some critical temperature.  However, the symmetry breaking mechanism in this case is the density of excitons in a given region which are introduced into the system via the external pumping source. 
By populating the vibron mode, which does not contribute directly 
to the lower polariton population, and allowing this mode to equilibrate and populate the exciton mode, it effectively {\em augments} the 
local exciton density such the the critical density can be reached at lower exciton pumping rates.  Increasing the temperature, 
causes as shift in the steady state populations of vibrons and excitons towards more population in the  higher-energy species.

\subsection{Rapid interconversion regime}

Central to our theory is the rate that free excitons are converted to lower polaritons.  To estimate this rate, let us assume that 
the ``golden rule'' is valid, such that the maximum value of the rate is given by
\begin{eqnarray}
r_{max} = \frac{2 \pi}{\hbar} | \Omega |^{2} \rho(k_{ex} = k_{cav})  \label{rstrong}
\end{eqnarray}
where $ \rho(k_{ex} = k_{cav})$ is the density of cavity states evaluated at the avoided crossing.   Since the cavity contains a two-dimensional gas of photons
with dispersion $E_{k} = \Delta  + \hbar \eta k^{2}$, the density of states is constant with 
\begin{eqnarray}
\rho(k_{ex} = k_{cav}) = \frac{\pi}{\hbar \eta} \label{dos}
\end{eqnarray}
Using the numerical values for our model cavity and 
setting $\Omega =0.05$eV, one obtains $r_{max} = 1.9\times 10^{7} {\rm ns}^{-1}$.  

We can follow a similar line of reasoning for the coupling between the exciton and vibron reservoirs. 
The vibron reservoir in our model corresponds to an ensemble of UP polaritons formed by 
mixing a cavity mode with a ground-electronic state vibrational mode of the molecules in the cavity.  
If we assume that the golden rule holds, then we can estimate the exciton to vibron conversion rate
by multiplying Eq.~\ref{rstrong} a Franck-Condon factor. {\rm i.e.}
$k = f_{0-1}^{2} \times r_{max}$.  Based upon arguments above, the exciton to vibron conversion rate
is at least an order of magnitude slower than the exciton to LP conversion. 

For the LP decay, let us assume that the cavity has a quality factor, $Q = 6000$, that corresponds to the
number of times a photon will traverse the cavity before escaping.  
Taking the cavity width to be $\approx$ 250 nm, one obtains an LP decay rate of $\gamma = 120 {\rm ns}^{-1}$. 

Since the conversion rates in this regime are far greater than in the slow conversion regime we just examined, the critical 
pumping strengths will be considerably lower.  However, the life times and dynamics will be considerably faster. 
In thinking about possible experimental situations, we consider the effect of a single 30 fs gaussian shaped pulse 
that drives the creation of an exciton gas. The density of excitons created will depend upon the intensity (and duration) of the pulse.  
 Prior to the arrival of the pulse, we seed the condensate with a very small amplitude  fluctuation at $x = 0$ which will have decayed 
 somewhat by the time the pulse arrives.

 \begin{figure}
\includegraphics[width=\columnwidth]{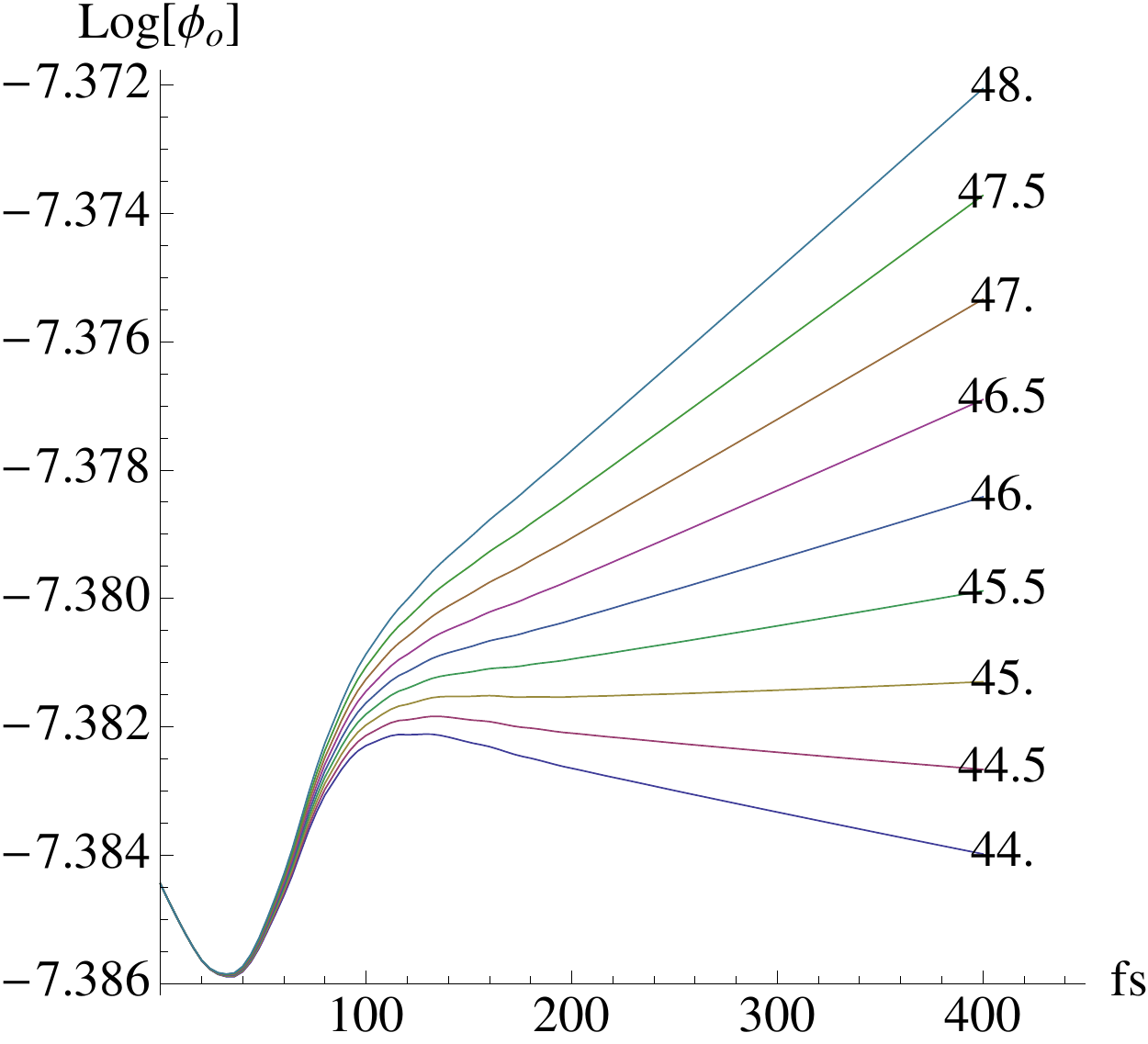}
\caption{Condensate amplitude at $x =0$  for a non-resonant cavity in the ``rapid-conversion'' model 
following a 30fs pulse with peak intensity of 60 ns$^{-1}$.  Each curve is labeled by the lattice temperature (in K).
}\label{pulse-amp}
\end{figure}
 
 In Fig.~\ref{pulse-amp} we show how the condensate population at the center of the simulation cell varies in time for a non-resonant 
 cavity system with $\Delta = 2.88 eV$ and at a fixed pumping strength of $p_{o} = 60 {\rm ns}^{-1}$.   
 At $t = 0fs$, a condensate fluctuation is introduced and at $t =60$fs  the
 30 fs excitation pulse reaches its maximum intensity and creates a population of excitons, which go on to produce
 more condensate as well as vibrons.  After 100 fs, the system evolves without further pumping and the population in the 
 condensate will either decay due to cavity loss, or continue to grow as excitons are converted to LP condensate. 
The various curves shown here correspond to different lattice temperatures.  For the case at hand, when the lattice 
temperature is above 45K, the condensate population continues to grow following the initial pulse.  This indicates that
sufficient exciton density is present to spawn the formation of a long-lived polariton condensate.  Below this temperature, 
the LP population simply decays.

\begin{figure}
\includegraphics[width=\columnwidth]{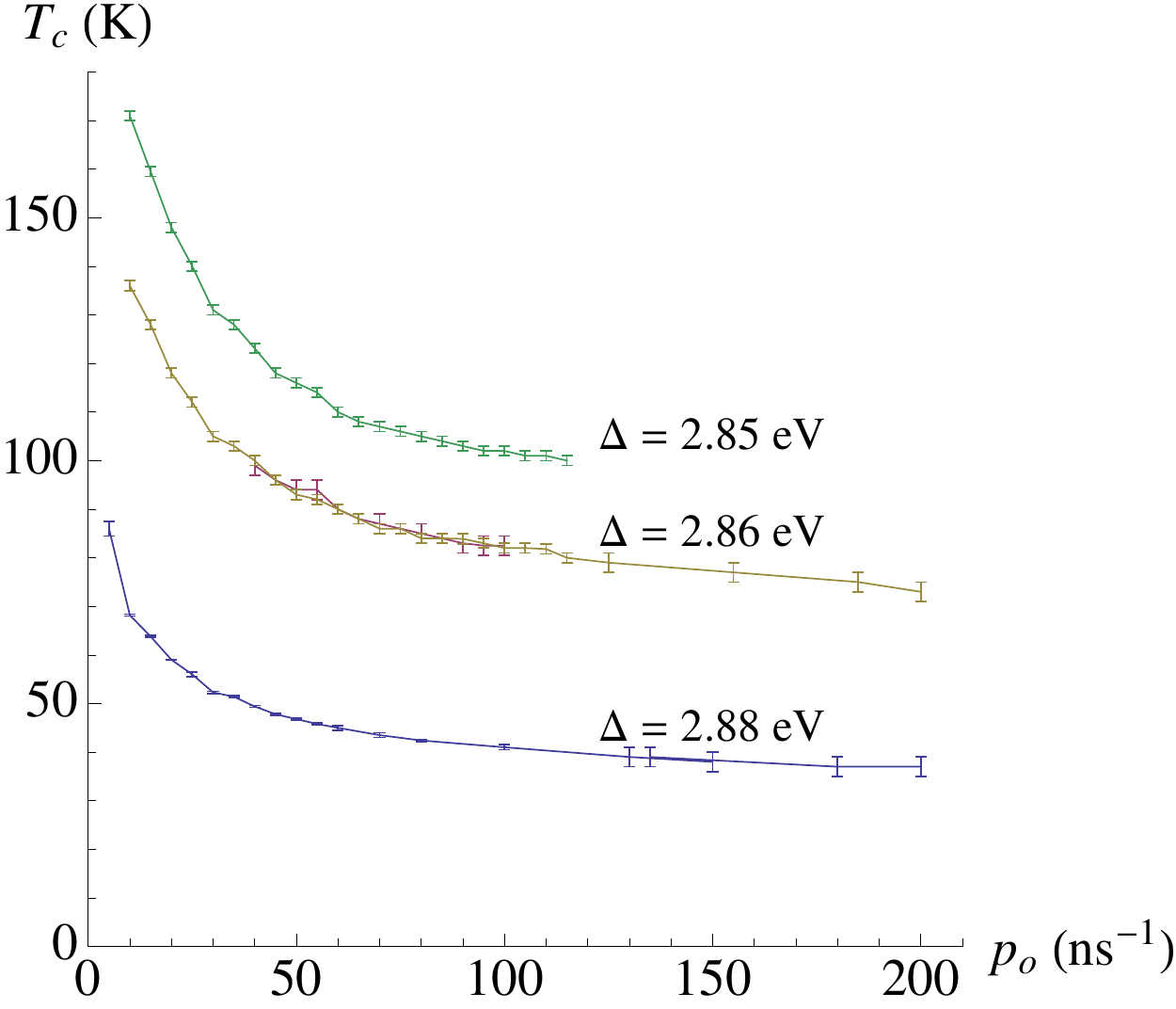}
\caption{Phase diagrams for non-resonant cavities in the ``rapid-conversion'' model.}\label{pulse-phased}
\end{figure}

Fig.~\ref{pulse-phased} shows a phase diagram for the off-resonant cavity 
in terms of the threshold temperature needed to produce a long-lived condensate at a given pumping strength.  
As temperature is decreased, the role of the vibronic reservoir is diminished and higher pumping strengths are 
required to from a long-lived condensate.  

\begin{figure}
%\subfigure[]{
\includegraphics[width=\columnwidth]{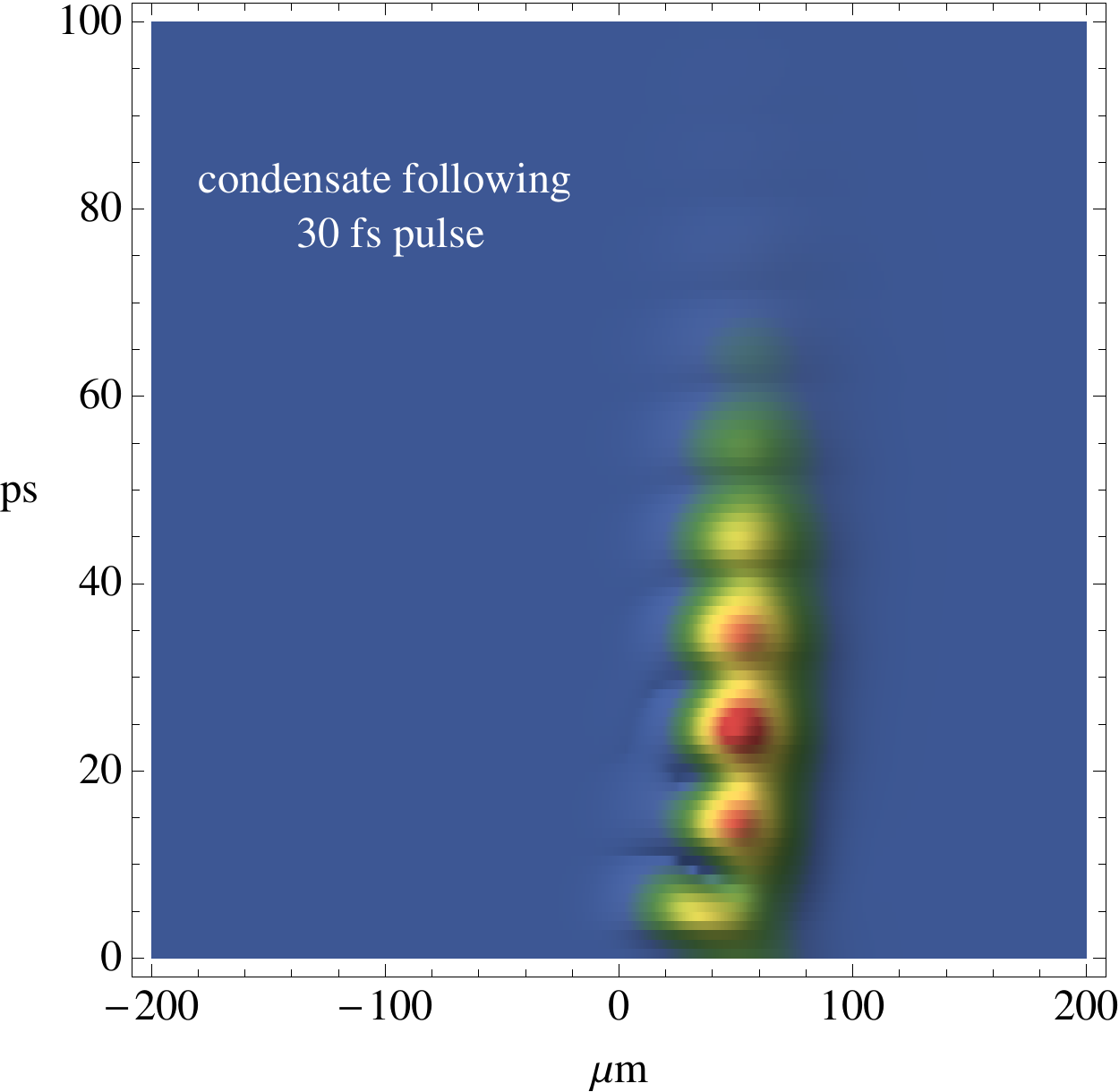}
%}
%\subfigure[]{\includegraphics[width=\columnwidth]{momentum}}
\caption{Time evolution of condensate in the ``rapid conversion'' model ($\Delta = 2.88$eV, $p_{o} = 60{\rm ns}^{-1}$). 
In this case, the seed-fluctuation was introduced about $x = 50 \mu{\rm m}$.}\label{longtime}
\end{figure}

 In Fig.~\ref{longtime}  we considered what happens if there is a spacial off-set between the pumping pulse and the initial 
condensate ``seed'' fluctuation.  We initialized the condensate with a small gaussian 
centered off-center at 50$\mu m$ and created  exciton density centered about $x = 0$ at 60 fs later.   
In this case, the condensate appears to migrate before decaying and there are a number of ``beats'' in the 
density itself.  The beats occur even when the seed fluctuation is located at $x = 0$ and are 
 due to  cyclic growth/decay kinetics between the condensate and exciton densities, much like what is observed 
 in predator/prey and disease models rather than to coherent motion of the condensate itself.  
At long times following the pulse, the condensate eventually decays due to cavity loss.   
It should be pointed out, however, that while the requisite pumping intensity in this model is lower than in the ``slow conversion'' case, 
the integrated density of the condensate wave function is considerably less.  
However, this density can be used to further seed the production of additional condensate if subsequent pulses 
are introduced.  

  \begin{figure}[b]
 \subfigure[]{\includegraphics[width=0.48\columnwidth]{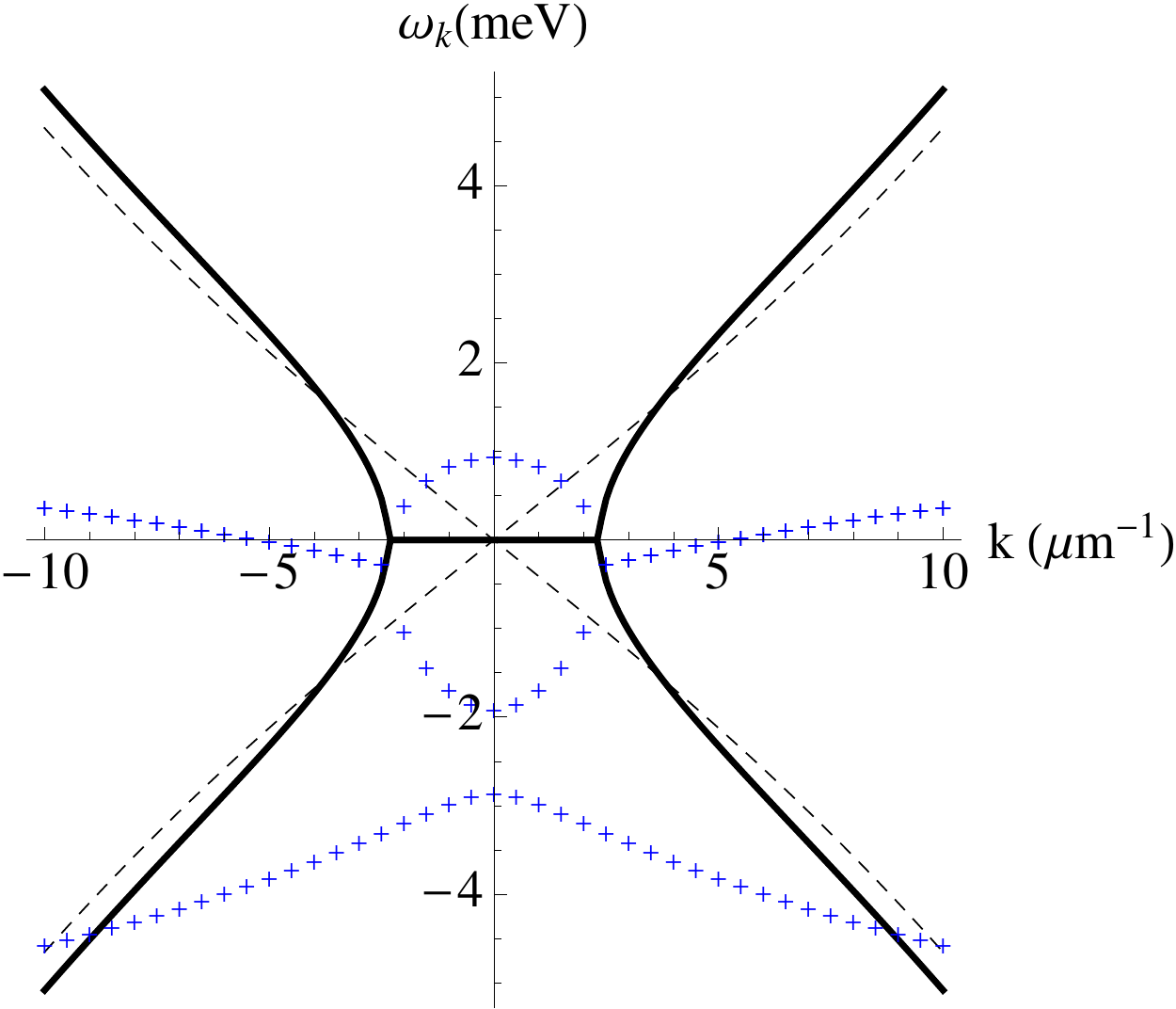}}
 \subfigure[]{\includegraphics[width=0.48\columnwidth]{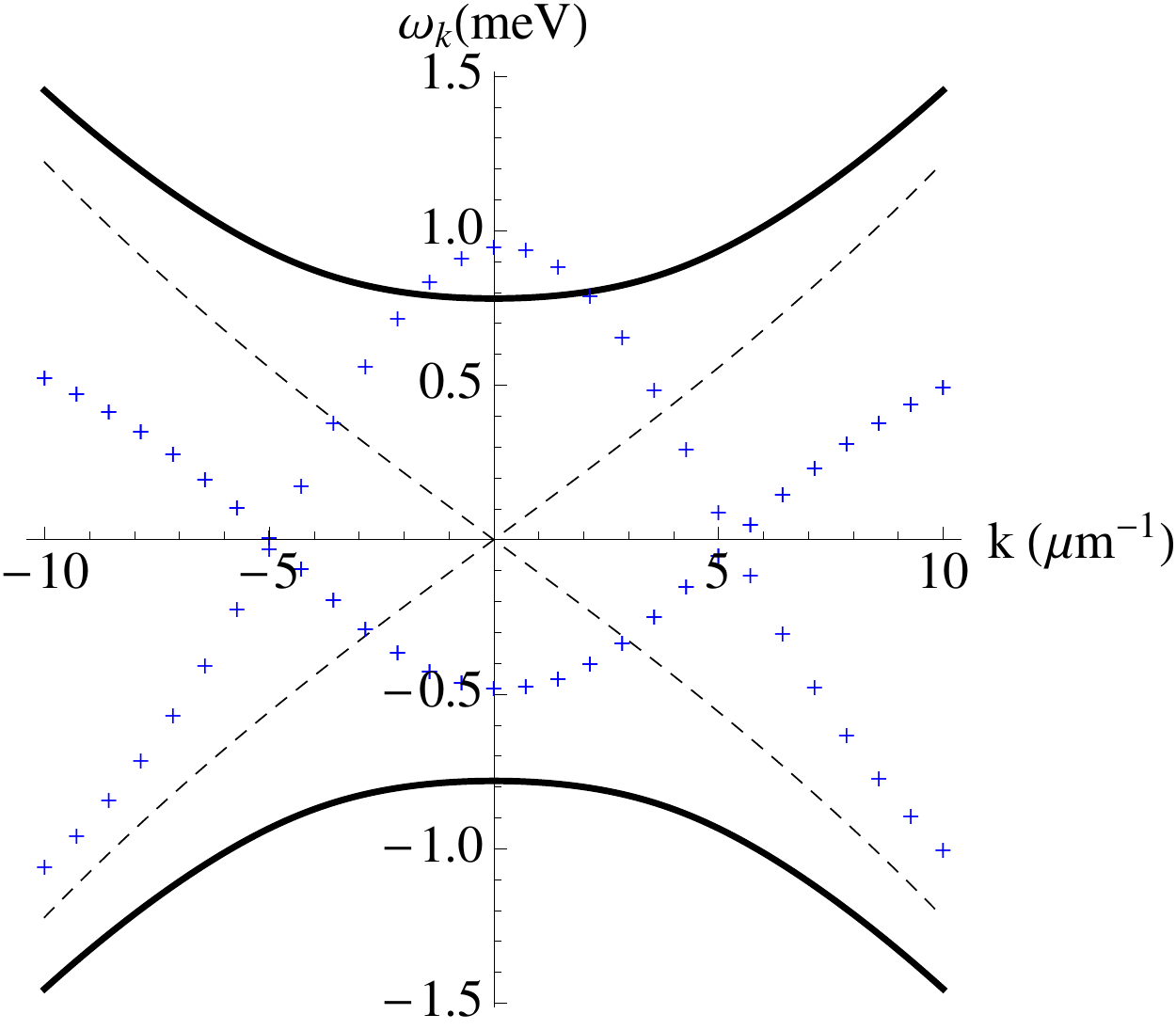}}
 \caption{ Real (line) and imaginary (+) components of the quasiparticle excitation spectrum for in the slow (a: $\gamma_{S}' \gg \gamma$)   
 and rapid (b: $\gamma_{S}' \ll \gamma $) conversion models.   Dashed curves are the Bogoliubov quasi-particle spectra for the equilibrium Bose condensate.   Unless otherwise specified, 
 we have set,  $g=1$,  $m_{LP} = 10^{-5}m_{e}$,   $\phi_{o}=1$  and $rS_{o}= 3$.  
  }\label{stab}
  \end{figure}

\section{Stability analysis}

We can introduce a time-dependent
fluctuation about the steady-state $\Psi_{o}= \{\phi_{o},\psi_{o}^{*}, S_{o},v_{o}\}$ 
by writing the solution to Eqs. 9 and 10  as $\Psi(0) = \Psi_{o} + \delta\Psi(t)$

Let us consider the time evolution of Eqs. 9 and 10  by writing $\Psi(t) = \{\phi(t),\psi^{*}(t), S(t),v(t)\}$
and introducing a fluctuation about its steady state such that $\Psi(t) = \Psi_{o} + \delta \Psi(t)$.
Introducing this into Eq. 9. and 10,  keeping only terms linear in $\delta\Psi$, and 
transforming from real to k-space, 
produces a set of coupled equations for the fluctuations:
\begin{eqnarray}
i\frac{\partial}{\partial t}
\left(
\begin{array}{c}
\delta\phi_{{\bf k}} \\ \delta\phi^{*}_{{\bf k} }\\ \delta S_{{\bf k}} \\ \delta v_{{\bf k}}
\end{array}
\right)
= {\cal M}({\bf k}) \left(
\begin{array}{c} 
\delta\phi_{{\bf k}} \\ \delta\phi^{*}_{{\bf k}} \\ \delta S_{{\bf k}} \\ \delta v_{{\bf k}}
\end{array}
\right),
\end{eqnarray}
where the stability matrix is given by
\begin{widetext}
\begin{eqnarray}
{\cal M}({\bf k})= 
\left[
\begin{array}{cccc}
\eta {\bf k}^{2} + (g |\phi_{o}|^{2} + \frac{i}{2}(r S_{o} - \gamma)) &   g(\phi_{o})^{2} &  \frac{i}{2}r \phi_{o}   & 0 \\
  -g(\phi_{o}^{*})^{2} &-\eta {\bf k}^{2}  - (g |\phi_{o}|^{2} - \frac{i}{2}(r S_{o} - \gamma)) &  \frac{i}{2}r \phi_{o}^{*}  & 0 \\
-i \phi_{o}^{*}S_{o}     & -i \phi_{o}S_{o}      &  -i (\gamma_{S}  +k  + r |\phi_{o}|^{2})  &  ik'\\
       0                      &       0                       &       i k                                          & - ik' 
\end{array}
\right].
\end{eqnarray}
 \end{widetext}
 In writing this, we assume the steady-state solutions are homogeneous and that wave-vector, ${\bf k}$, remains a constant of the motion.
 We have also used $\mu = g |\phi_{o}|^{2}$ as the chemical potential of the condensate.
( Note that the $k$ and $k'$ appearing in the matrix elements coupling the reservoirs are rate constants and not wave-vectors.)
The stability matrix, ${\cal M}$,  given above is similar to the equations of motion
 reported by Wouters  and Carusotto\cite{PhysRevLett.99.140402}
 and by Byrnes {\em et al.}\cite{PhysRevB.85.075130}
 In the absence of the reservoir, the eigenvalues of the upper $2\times 2$ block of ${\cal M}({\bf k})$ are the Bogoliubov modes 
 corresponding to excitations of the  condensate.
 These modes vanish at ${\bf k}=0$
 corresponding to a Goldstone brach that can be understood as a slow rotation of the condensate across the sample\cite{PhysRevLett.99.140402} 
  and the linear dispersion about ${\bf k}=0$ is characteristic of a superfluid state.

 Let us consider the stability of the polariton/exciton/vibron equations in the steady-state limit
by first assuming that the exciton and vibron reservoirs are equilibrated rapidly.  This allows us to 
write an effective exciton decay constant, $\gamma'_{S}$, that will depend both upon the 
lattice temperature and the cavity offset.  
  Including the coupling to the reservoir, two regimes can be identified. 
 First, in the limit where $\gamma'_{S} \gg \gamma$ in which the reservoir relaxation is very rapid compared to the polariton relaxation rate, 
 the dispersion around ${\bf k}=0$ is flat corresponding to a diffusive Goldstone mode which is in stark contrast to the linear dispersion of the
 sound-mode in an equilibrium Bose condensate.   
% The two limits are akin to the cross-over between BEC and a BCS superconductor.\cite{Gaebler:2010fk}
  For a Bose condensate, the 
 gap at ${\bf k}=0$  necessarily vanishes and the dispersion about ${\bf k}=0$ is linear corresponding to the formation of a 
 superfluid with the slope proportional to the sound velocity.
 In the non-equilibrium case, the sound velocity vanishes as well as seen in Fig.~\ref{stab}a 
  which  corresponds to our ``slow''  regime in which the exciton gas decays rapidly compared to the 
  polariton condensate. 
  
 On the other hand, if $\gamma \gg \gamma'_{S}$ a gap opens at ${\bf k}=0$ and the polaritons again take on an effective mass. 
 The gap and the flattening of the spectrum is akin to the blue-shift observed by Kasprzak for a polariton condensate  driven well above the condensation threshold\cite{Kasprzak:2006jt,Kasprzak:2006mb,Kasprzak:2008lh,Kasprzak:2008mi,Richard:2005tw}.  It was speculated that the flattening could be due to either polariton localization in 
 real-space or to pump and decay processes. 
Here, as in Refs. \onlinecite{PhysRevLett.96.066405,PhysRevLett.96.230602} the flattening and gap is entirely due to
competing decay processes between the exciton reservoir and the polariton gas.

% is similar to what 
%is seen in the BCS model of superconductivity In fact, our model produces
%dispersions very similar to those reported in Gaebler, {\em et al.} \cite{Gaebler:2010fk} for the BEC/BCS crossover in 
%a strongly interacting fermion gas. 
%As exhibited in Fig.~\ref{stab}d in which we use parameters corresponding to the
% ``rapid conversion'' model, we are clearly in the BCS regime.
% Since the lifetime of the exciton gas depends upon both the natural 
%  lifetime of the exciton and the rate of interconversion with the secondary vibron reservoir, we speculate that 
%  it may be   possible to tune the system between the BEC and BCS regimes by manipulating the cavity. 
%  

 \section{Discussion}

In this paper we have focused upon the role that molecular vibrational excitations may play in the formation of a LP condensate in a
microcavity containing a thin-film organic semiconductor, such as anthracene.   In our model, molecular vibrational excitations coupled to the photon field serve as a secondary thermal reservoir for free excitons.  
Thermal fluctuations from the vibron reservoir can augment the exciton density 
such that at finite lattice temperature polariton condensation can be achieved at lower exciton pumping rates. 
We assumed that the vibron reservoir involved only a single vibrational model per molecule.  In a realistic 
system, there will be multiple vibrational modes as evidenced by the various Franck-Condon peaks in the emission spectrum of 
many organic semiconductors.  Assuming the mechanism presented here holds true in the multi-mode case, these Franck-Condon modes 
should provide a reservoir for the exciton gas and greatly facilitate the formation of a stable LP condensate. 

We explored two limits of the model. In the ``slow conversion'' model, we assumed that the rate 
of conversion between continuously driven  and replenished
gas of free excitons and a vibronic reservoir level was comparable to the rate of conversion between the exciton and a lower polariton 
condensate.  In this limit, the temperature of the molecular crystal lattice plays an 
important role in providing a sufficient exciton density
to spawn the growth of fluctuations about the condensate vacuum.  At long times, 
this model produces steady state solutions with populations in 
both condensate and non-condensate channels.  
In the ``rapid conversion'' limit, we assumed that the conversion of the exciton gas to
 LP condensate could be estimated from the Rabi splitting between the 
upper and lower polariton branches and the density of states of photons in a two-dimensional cavity.  In this model, the exciton gas is pumped impulsively via a single 30 fs pulse.  Here, too, the growth or decay of condensate fluctuations hinges upon the pumping  being above a threshold intensity and shows a pronounced sensitivity to the temperature of the lattice when the cavity is off-resonance with the 
exciton. In both models, it is highly encouraging that the threshold intensities are well within the margins of current experimental capabilities.

Lastly, along a similar vein, one can model an electrically pumped system in which electrons and holes are injected onto a microcavity device and recombine to form singlet and triplet excitons. We suspect that both the triplet excitons and bound charge-transfer contact pairs may
serve a similar role as the vibronic reservoir in the present model.   We are currently exploring this possibility. 

\acknowledgments

The work at the University of Houston was funded in part by the National Science Foundation (CHE-1011894) and the Robert A. Welch Foundation (E-1334). 
CS acknowledges support from the Canada Research Chair in Organic Semiconductor Materials.


\begin{thebibliography}{36}
\expandafter\ifx\csname natexlab\endcsname\relax\def\natexlab#1{#1}\fi
\expandafter\ifx\csname bibnamefont\endcsname\relax
  \def\bibnamefont#1{#1}\fi
\expandafter\ifx\csname bibfnamefont\endcsname\relax
  \def\bibfnamefont#1{#1}\fi
\expandafter\ifx\csname citenamefont\endcsname\relax
  \def\citenamefont#1{#1}\fi
\expandafter\ifx\csname url\endcsname\relax
  \def\url#1{\texttt{#1}}\fi
\expandafter\ifx\csname urlprefix\endcsname\relax\def\urlprefix{URL }\fi
\providecommand{\bibinfo}[2]{#2}
\providecommand{\eprint}[2][]{\url{#2}}

\bibitem[{\citenamefont{Kasprzak
  et~al.}(2006{\natexlab{a}})\citenamefont{Kasprzak, Richard, Kundermann, Baas,
  Jeambriun, Keeling, Marchetti, Szymanska, Andr\'{e}, Staehli
  et~al.}}]{Kasprzak:2006mz}
\bibinfo{author}{\bibfnamefont{J.}~\bibnamefont{Kasprzak}},
  \bibinfo{author}{\bibfnamefont{M.}~\bibnamefont{Richard}},
  \bibinfo{author}{\bibfnamefont{S.}~\bibnamefont{Kundermann}},
  \bibinfo{author}{\bibfnamefont{A.}~\bibnamefont{Baas}},
  \bibinfo{author}{\bibfnamefont{P.}~\bibnamefont{Jeambriun}},
  \bibinfo{author}{\bibfnamefont{J.}~\bibnamefont{Keeling}},
  \bibinfo{author}{\bibfnamefont{F.~M.} \bibnamefont{Marchetti}},
  \bibinfo{author}{\bibfnamefont{M.~H.} \bibnamefont{Szymanska}},
  \bibinfo{author}{\bibfnamefont{R.}~\bibnamefont{Andr\'{e}}},
  \bibinfo{author}{\bibfnamefont{J.~L.} \bibnamefont{Staehli}},
  \bibnamefont{et~al.}, \bibinfo{journal}{Nature}
  \textbf{\bibinfo{volume}{443}}, \bibinfo{pages}{28}
  (\bibinfo{year}{2006}{\natexlab{a}}).

\bibitem[{\citenamefont{Malpuech et~al.}(2009)\citenamefont{Malpuech,
  Solnyshkov, and Shelykh}}]{malpuech:21}
\bibinfo{author}{\bibfnamefont{G.}~\bibnamefont{Malpuech}},
  \bibinfo{author}{\bibfnamefont{D.}~\bibnamefont{Solnyshkov}},
  \bibnamefont{and} \bibinfo{author}{\bibfnamefont{I.}~\bibnamefont{Shelykh}},
  in \emph{\bibinfo{booktitle}{AIP Conference Proceedings}}, edited by
  \bibinfo{editor}{\bibfnamefont{D.~N.} \bibnamefont{Chigrin}}
  (\bibinfo{publisher}{American Institute of Physics}, \bibinfo{year}{2009}),
  vol. \bibinfo{volume}{1176}, pp. \bibinfo{pages}{21--22},
  \urlprefix\url{http://link.aip.org/link/?APC/1176/21/1}.

\bibitem[{\citenamefont{Malpuech et~al.}(2007)\citenamefont{Malpuech,
  Solnyshkov, Ouerdane, Glazov, and Shelykh}}]{Malpuech:2007ec}
\bibinfo{author}{\bibfnamefont{G.}~\bibnamefont{Malpuech}},
  \bibinfo{author}{\bibfnamefont{D.~D.} \bibnamefont{Solnyshkov}},
  \bibinfo{author}{\bibfnamefont{H.}~\bibnamefont{Ouerdane}},
  \bibinfo{author}{\bibfnamefont{M.~M.} \bibnamefont{Glazov}},
  \bibnamefont{and} \bibinfo{author}{\bibfnamefont{I.}~\bibnamefont{Shelykh}},
  \bibinfo{journal}{Phys. Rev. Lett.} \textbf{\bibinfo{volume}{98}}
  (\bibinfo{year}{2007}),
  \urlprefix\url{http://dx.doi.org/10.1103/PhysRevLett.98.206402}.

\bibitem[{\citenamefont{Agranovich et~al.}(2003)\citenamefont{Agranovich,
  Litinskaia, and Lidzey}}]{PhysRevB.67.085311}
\bibinfo{author}{\bibfnamefont{V.~M.} \bibnamefont{Agranovich}},
  \bibinfo{author}{\bibfnamefont{M.}~\bibnamefont{Litinskaia}},
  \bibnamefont{and} \bibinfo{author}{\bibfnamefont{D.~G.}
  \bibnamefont{Lidzey}}, \bibinfo{journal}{Phys. Rev. B}
  \textbf{\bibinfo{volume}{67}}, \bibinfo{pages}{085311}
  (\bibinfo{year}{2003}).

\bibitem[{\citenamefont{Carusotto and Ciuti}(2005)}]{PhysRevB.72.125335}
\bibinfo{author}{\bibfnamefont{I.}~\bibnamefont{Carusotto}} \bibnamefont{and}
  \bibinfo{author}{\bibfnamefont{C.}~\bibnamefont{Ciuti}},
  \bibinfo{journal}{Phys. Rev. B} \textbf{\bibinfo{volume}{72}},
  \bibinfo{pages}{125335} (\bibinfo{year}{2005}).

\bibitem[{\citenamefont{Maragkou et~al.}(2010)\citenamefont{Maragkou, Grundy,
  Wertz, Lema\^\i{}tre, Sagnes, Senellart, Bloch, and
  Lagoudakis}}]{PhysRevB.81.081307}
\bibinfo{author}{\bibfnamefont{M.}~\bibnamefont{Maragkou}},
  \bibinfo{author}{\bibfnamefont{A.~J.~D.} \bibnamefont{Grundy}},
  \bibinfo{author}{\bibfnamefont{E.}~\bibnamefont{Wertz}},
  \bibinfo{author}{\bibfnamefont{A.}~\bibnamefont{Lema\^\i{}tre}},
  \bibinfo{author}{\bibfnamefont{I.}~\bibnamefont{Sagnes}},
  \bibinfo{author}{\bibfnamefont{P.}~\bibnamefont{Senellart}},
  \bibinfo{author}{\bibfnamefont{J.}~\bibnamefont{Bloch}}, \bibnamefont{and}
  \bibinfo{author}{\bibfnamefont{P.~G.} \bibnamefont{Lagoudakis}},
  \bibinfo{journal}{Phys. Rev. B} \textbf{\bibinfo{volume}{81}},
  \bibinfo{pages}{081307} (\bibinfo{year}{2010}).

\bibitem[{\citenamefont{Deng et~al.}(2010)\citenamefont{Deng, Haug, and
  Yamamoto}}]{RevModPhys.82.1489}
\bibinfo{author}{\bibfnamefont{H.}~\bibnamefont{Deng}},
  \bibinfo{author}{\bibfnamefont{H.}~\bibnamefont{Haug}}, \bibnamefont{and}
  \bibinfo{author}{\bibfnamefont{Y.}~\bibnamefont{Yamamoto}},
  \bibinfo{journal}{Rev. Mod. Phys.} \textbf{\bibinfo{volume}{82}},
  \bibinfo{pages}{1489} (\bibinfo{year}{2010}).

\bibitem[{\citenamefont{Deng et~al.}(2002)\citenamefont{Deng, Weihs, Santori,
  Bloch, and Yamamoto}}]{Deng:2002zt}
\bibinfo{author}{\bibfnamefont{H.}~\bibnamefont{Deng}},
  \bibinfo{author}{\bibfnamefont{G.}~\bibnamefont{Weihs}},
  \bibinfo{author}{\bibfnamefont{C.}~\bibnamefont{Santori}},
  \bibinfo{author}{\bibfnamefont{J.}~\bibnamefont{Bloch}}, \bibnamefont{and}
  \bibinfo{author}{\bibfnamefont{Y.}~\bibnamefont{Yamamoto}},
  \bibinfo{journal}{Science} \textbf{\bibinfo{volume}{298}},
  \bibinfo{pages}{199} (\bibinfo{year}{2002}).

\bibitem[{\citenamefont{Snoke}(2002)}]{DavidSnoke11152002}
\bibinfo{author}{\bibfnamefont{D.}~\bibnamefont{Snoke}},
  \bibinfo{journal}{Science} \textbf{\bibinfo{volume}{298}},
  \bibinfo{pages}{1368} (\bibinfo{year}{2002}),
  \eprint{http://www.sciencemag.org/cgi/reprint/298/5597/1368.pdf},
  \urlprefix\url{http://www.sciencemag.org/cgi/content/abstract/298/5597/1368}.

\bibitem[{\citenamefont{Utsunomiya et~al.}(2008)\citenamefont{Utsunomiya, Tian,
  Roumpos, Lai, Kumada, Fujisawa, Kuwata-Gonokami, Loffler, Hofling, Forchel
  et~al.}}]{Utsunomiya:2008fk}
\bibinfo{author}{\bibfnamefont{S.}~\bibnamefont{Utsunomiya}},
  \bibinfo{author}{\bibfnamefont{L.}~\bibnamefont{Tian}},
  \bibinfo{author}{\bibfnamefont{G.}~\bibnamefont{Roumpos}},
  \bibinfo{author}{\bibfnamefont{C.~W.} \bibnamefont{Lai}},
  \bibinfo{author}{\bibfnamefont{N.}~\bibnamefont{Kumada}},
  \bibinfo{author}{\bibfnamefont{T.}~\bibnamefont{Fujisawa}},
  \bibinfo{author}{\bibfnamefont{M.}~\bibnamefont{Kuwata-Gonokami}},
  \bibinfo{author}{\bibfnamefont{A.}~\bibnamefont{Loffler}},
  \bibinfo{author}{\bibfnamefont{S.}~\bibnamefont{Hofling}},
  \bibinfo{author}{\bibfnamefont{A.}~\bibnamefont{Forchel}},
  \bibnamefont{et~al.}, \bibinfo{journal}{Nat Phys}
  \textbf{\bibinfo{volume}{4}}, \bibinfo{pages}{700} (\bibinfo{year}{2008}),
  \urlprefix\url{http://dx.doi.org/10.1038/nphys1034}.

\bibitem[{\citenamefont{Houdr\'e et~al.}(2000)\citenamefont{Houdr\'e, Weisbuch,
  Stanley, Oesterle, and Ilegems}}]{PhysRevLett.85.2793}
\bibinfo{author}{\bibfnamefont{R.}~\bibnamefont{Houdr\'e}},
  \bibinfo{author}{\bibfnamefont{C.}~\bibnamefont{Weisbuch}},
  \bibinfo{author}{\bibfnamefont{R.~P.} \bibnamefont{Stanley}},
  \bibinfo{author}{\bibfnamefont{U.}~\bibnamefont{Oesterle}}, \bibnamefont{and}
  \bibinfo{author}{\bibfnamefont{M.}~\bibnamefont{Ilegems}},
  \bibinfo{journal}{Phys. Rev. Lett.} \textbf{\bibinfo{volume}{85}},
  \bibinfo{pages}{2793} (\bibinfo{year}{2000}),
  \urlprefix\url{http://link.aps.org/doi/10.1103/PhysRevLett.85.2793}.

\bibitem[{\citenamefont{Chovan et~al.}(2008)\citenamefont{Chovan, Perakis,
  Ceccarelli, and Lidzey}}]{chovan:045320}
\bibinfo{author}{\bibfnamefont{J.}~\bibnamefont{Chovan}},
  \bibinfo{author}{\bibfnamefont{I.~E.} \bibnamefont{Perakis}},
  \bibinfo{author}{\bibfnamefont{S.}~\bibnamefont{Ceccarelli}},
  \bibnamefont{and} \bibinfo{author}{\bibfnamefont{D.~G.}
  \bibnamefont{Lidzey}}, \bibinfo{journal}{Physical Review B (Condensed Matter
  and Materials Physics)} \textbf{\bibinfo{volume}{78}}, \bibinfo{eid}{045320}
  (pages~\bibinfo{numpages}{5}) (\bibinfo{year}{2008}),
  \urlprefix\url{http://link.aps.org/abstract/PRB/v78/e045320}.

\bibitem[{\citenamefont{Lidzey et~al.}(2002)\citenamefont{Lidzey, Fox, Rahn,
  Skolnick, Agranovich, and Walker}}]{PhysRevB.65.195312}
\bibinfo{author}{\bibfnamefont{D.~G.} \bibnamefont{Lidzey}},
  \bibinfo{author}{\bibfnamefont{A.~M.} \bibnamefont{Fox}},
  \bibinfo{author}{\bibfnamefont{M.~D.} \bibnamefont{Rahn}},
  \bibinfo{author}{\bibfnamefont{M.~S.} \bibnamefont{Skolnick}},
  \bibinfo{author}{\bibfnamefont{V.~M.} \bibnamefont{Agranovich}},
  \bibnamefont{and} \bibinfo{author}{\bibfnamefont{S.}~\bibnamefont{Walker}},
  \bibinfo{journal}{Phys. Rev. B} \textbf{\bibinfo{volume}{65}},
  \bibinfo{pages}{195312} (\bibinfo{year}{2002}).

\bibitem[{\citenamefont{Lidzey et~al.}(1999)\citenamefont{Lidzey, Bradley,
  Virgili, Armitage, Skolnick, and Walker}}]{PhysRevLett.82.3316}
\bibinfo{author}{\bibfnamefont{D.~G.} \bibnamefont{Lidzey}},
  \bibinfo{author}{\bibfnamefont{D.~D.~C.} \bibnamefont{Bradley}},
  \bibinfo{author}{\bibfnamefont{T.}~\bibnamefont{Virgili}},
  \bibinfo{author}{\bibfnamefont{A.}~\bibnamefont{Armitage}},
  \bibinfo{author}{\bibfnamefont{M.~S.} \bibnamefont{Skolnick}},
  \bibnamefont{and} \bibinfo{author}{\bibfnamefont{S.}~\bibnamefont{Walker}},
  \bibinfo{journal}{Phys. Rev. Lett.} \textbf{\bibinfo{volume}{82}},
  \bibinfo{pages}{3316} (\bibinfo{year}{1999}).

\bibitem[{\citenamefont{Somaschi et~al.}(2011)\citenamefont{Somaschi,
  Mouchliadis, Coles, Perakis, Lidzey, Lagoudakis, and
  Savvidis}}]{somaschi:143303}
\bibinfo{author}{\bibfnamefont{N.}~\bibnamefont{Somaschi}},
  \bibinfo{author}{\bibfnamefont{L.}~\bibnamefont{Mouchliadis}},
  \bibinfo{author}{\bibfnamefont{D.}~\bibnamefont{Coles}},
  \bibinfo{author}{\bibfnamefont{I.~E.} \bibnamefont{Perakis}},
  \bibinfo{author}{\bibfnamefont{D.~G.} \bibnamefont{Lidzey}},
  \bibinfo{author}{\bibfnamefont{P.~G.} \bibnamefont{Lagoudakis}},
  \bibnamefont{and} \bibinfo{author}{\bibfnamefont{P.~G.}
  \bibnamefont{Savvidis}}, \bibinfo{journal}{Applied Physics Letters}
  \textbf{\bibinfo{volume}{99}}, \bibinfo{eid}{143303}
  (pages~\bibinfo{numpages}{3}) (\bibinfo{year}{2011}),
  \urlprefix\url{http://link.aip.org/link/?APL/99/143303/1}.

\bibitem[{\citenamefont{Coles et~al.}(2011)\citenamefont{Coles, Michetti,
  Clark, Tsoi, Adawi, Kim, and Lidzey}}]{lidzey:2011}
\bibinfo{author}{\bibfnamefont{D.~M.} \bibnamefont{Coles}},
  \bibinfo{author}{\bibfnamefont{P.}~\bibnamefont{Michetti}},
  \bibinfo{author}{\bibfnamefont{C.}~\bibnamefont{Clark}},
  \bibinfo{author}{\bibfnamefont{W.~C.} \bibnamefont{Tsoi}},
  \bibinfo{author}{\bibfnamefont{A.~M.} \bibnamefont{Adawi}},
  \bibinfo{author}{\bibfnamefont{J.-S.} \bibnamefont{Kim}}, \bibnamefont{and}
  \bibinfo{author}{\bibfnamefont{D.~G.} \bibnamefont{Lidzey}},
  \bibinfo{journal}{Advanced Functional Materials}
  \textbf{\bibinfo{volume}{21}}, \bibinfo{pages}{3691} (\bibinfo{year}{2011}),
  ISSN \bibinfo{issn}{1616-3028},
  \urlprefix\url{http://dx.doi.org/10.1002/adfm.201100756}.

\bibitem[{\citenamefont{Bittner and Silva}(2012)}]{BittnerJCP2012b}
\bibinfo{author}{\bibfnamefont{E.~R.} \bibnamefont{Bittner}} \bibnamefont{and}
  \bibinfo{author}{\bibfnamefont{C.}~\bibnamefont{Silva}}, \bibinfo{journal}{J.
  Chem. Phys.-in press}  (\bibinfo{year}{2012}).

\bibitem[{\citenamefont{Bittner et~al.}(2012)\citenamefont{Bittner, Zaster, and
  Silva}}]{C2CP23204A}
\bibinfo{author}{\bibfnamefont{E.}~\bibnamefont{Bittner}},
  \bibinfo{author}{\bibfnamefont{S.}~\bibnamefont{Zaster}}, \bibnamefont{and}
  \bibinfo{author}{\bibfnamefont{C.}~\bibnamefont{Silva}},
  \bibinfo{journal}{Phys. Chem. Chem. Phys.} pp.~\bibinfo{pages}{--}
  (\bibinfo{year}{2012}), \urlprefix\url{http://dx.doi.org/10.1039/C2CP23204A}.

\bibitem[{\citenamefont{Yamagata et~al.}(2011)\citenamefont{Yamagata, Norton,
  Hontz, Olivier, Beljonne, Br\'{e}das, Silbey, and Spano}}]{yamagata:204703}
\bibinfo{author}{\bibfnamefont{H.}~\bibnamefont{Yamagata}},
  \bibinfo{author}{\bibfnamefont{J.}~\bibnamefont{Norton}},
  \bibinfo{author}{\bibfnamefont{E.}~\bibnamefont{Hontz}},
  \bibinfo{author}{\bibfnamefont{Y.}~\bibnamefont{Olivier}},
  \bibinfo{author}{\bibfnamefont{D.}~\bibnamefont{Beljonne}},
  \bibinfo{author}{\bibfnamefont{J.~L.} \bibnamefont{Br\'{e}das}},
  \bibinfo{author}{\bibfnamefont{R.~J.} \bibnamefont{Silbey}},
  \bibnamefont{and} \bibinfo{author}{\bibfnamefont{F.~C.} \bibnamefont{Spano}},
  \bibinfo{journal}{The Journal of Chemical Physics}
  \textbf{\bibinfo{volume}{134}}, \bibinfo{eid}{204703}
  (pages~\bibinfo{numpages}{11}) (\bibinfo{year}{2011}),
  \urlprefix\url{http://link.aip.org/link/?JCP/134/204703/1}.

\bibitem[{\citenamefont{Bree and Lyons}(1960)}]{JR9600005206}
\bibinfo{author}{\bibfnamefont{A.}~\bibnamefont{Bree}} \bibnamefont{and}
  \bibinfo{author}{\bibfnamefont{L.~E.} \bibnamefont{Lyons}},
  \bibinfo{journal}{J. Chem. Soc.} pp. \bibinfo{pages}{5206--5212}
  (\bibinfo{year}{1960}),
  \urlprefix\url{http://dx.doi.org/10.1039/JR9600005206}.

\bibitem[{\citenamefont{K\'{e}na-Cohen and Forrest}(2008)}]{kena-cohen:073205}
\bibinfo{author}{\bibfnamefont{S.}~\bibnamefont{K\'{e}na-Cohen}}
  \bibnamefont{and} \bibinfo{author}{\bibfnamefont{S.~R.}
  \bibnamefont{Forrest}}, \bibinfo{journal}{Physical Review B (Condensed Matter
  and Materials Physics)} \textbf{\bibinfo{volume}{77}}, \bibinfo{eid}{073205}
  (pages~\bibinfo{numpages}{4}) (\bibinfo{year}{2008}),
  \urlprefix\url{http://link.aps.org/abstract/PRB/v77/e073205}.

\bibitem[{\citenamefont{Kena-Cohen and Forrest}(2010)}]{Kena-CohenS.:2010fk}
\bibinfo{author}{\bibfnamefont{S.}~\bibnamefont{Kena-Cohen}} \bibnamefont{and}
  \bibinfo{author}{\bibfnamefont{S.~R.} \bibnamefont{Forrest}},
  \bibinfo{journal}{Nature Photonics} \textbf{\bibinfo{volume}{4}},
  \bibinfo{pages}{371} (\bibinfo{year}{2010}),
  \urlprefix\url{http://dx.doi.org/10.1038/nphoton.2010.86}.

\bibitem[{\citenamefont{Holland}(1993)}]{Holland:1993}
\bibinfo{author}{\bibfnamefont{P.~R.} \bibnamefont{Holland}},
  \emph{\bibinfo{title}{The Quantum Theory of Motion: An Account of the de
  Broglie-Bohm Causal Interpretation of Quantum Mechanics}}
  (\bibinfo{publisher}{Cambridge University Press}, \bibinfo{year}{1993}).

\bibitem[{\citenamefont{Szyma\ifmmode~\acute{n}\else \'{n}\fi{}ska
  et~al.}(2006)\citenamefont{Szyma\ifmmode~\acute{n}\else \'{n}\fi{}ska,
  Keeling, and Littlewood}}]{PhysRevLett.96.230602}
\bibinfo{author}{\bibfnamefont{M.~H.} \bibnamefont{Szyma\ifmmode~\acute{n}\else
  \'{n}\fi{}ska}}, \bibinfo{author}{\bibfnamefont{J.}~\bibnamefont{Keeling}},
  \bibnamefont{and} \bibinfo{author}{\bibfnamefont{P.~B.}
  \bibnamefont{Littlewood}}, \bibinfo{journal}{Phys. Rev. Lett.}
  \textbf{\bibinfo{volume}{96}}, \bibinfo{pages}{230602}
  (\bibinfo{year}{2006}).

\bibitem[{\citenamefont{Keeling}(2011)}]{PhysRevLett.107.080402}
\bibinfo{author}{\bibfnamefont{J.}~\bibnamefont{Keeling}},
  \bibinfo{journal}{Phys. Rev. Lett.} \textbf{\bibinfo{volume}{107}},
  \bibinfo{pages}{080402} (\bibinfo{year}{2011}),
  \urlprefix\url{http://link.aps.org/doi/10.1103/PhysRevLett.107.080402}.

\bibitem[{\citenamefont{Wouters and Carusotto}(2007)}]{PhysRevLett.99.140402}
\bibinfo{author}{\bibfnamefont{M.}~\bibnamefont{Wouters}} \bibnamefont{and}
  \bibinfo{author}{\bibfnamefont{I.}~\bibnamefont{Carusotto}},
  \bibinfo{journal}{Phys. Rev. Lett.} \textbf{\bibinfo{volume}{99}},
  \bibinfo{pages}{140402} (\bibinfo{year}{2007}),
  \urlprefix\url{http://link.aps.org/doi/10.1103/PhysRevLett.99.140402}.

\bibitem[{MMa(2010)}]{MMa8.0}
\emph{\bibinfo{title}{Mathematica}} (\bibinfo{publisher}{{Wolfram Research,
  Inc.}}, \bibinfo{address}{Champaign, IL}, \bibinfo{year}{2010}),
  \bibinfo{note}{version 8}.

\bibitem[{\citenamefont{Salcedo et~al.}(1978)\citenamefont{Salcedo, Siegman,
  Dlott, and Fayer}}]{PhysRevLett.41.131}
\bibinfo{author}{\bibfnamefont{J.~R.} \bibnamefont{Salcedo}},
  \bibinfo{author}{\bibfnamefont{A.~E.} \bibnamefont{Siegman}},
  \bibinfo{author}{\bibfnamefont{D.~D.} \bibnamefont{Dlott}}, \bibnamefont{and}
  \bibinfo{author}{\bibfnamefont{M.~D.} \bibnamefont{Fayer}},
  \bibinfo{journal}{Phys. Rev. Lett.} \textbf{\bibinfo{volume}{41}},
  \bibinfo{pages}{131} (\bibinfo{year}{1978}),
  \urlprefix\url{http://link.aps.org/doi/10.1103/PhysRevLett.41.131}.

\bibitem[{\citenamefont{Kenkre and Schmid}(1985)}]{PhysRevB.31.2430}
\bibinfo{author}{\bibfnamefont{V.~M.} \bibnamefont{Kenkre}} \bibnamefont{and}
  \bibinfo{author}{\bibfnamefont{D.}~\bibnamefont{Schmid}},
  \bibinfo{journal}{Phys. Rev. B} \textbf{\bibinfo{volume}{31}},
  \bibinfo{pages}{2430} (\bibinfo{year}{1985}).

\bibitem[{\citenamefont{Byrnes et~al.}(2012)\citenamefont{Byrnes, Horikiri,
  Ishida, Fraser, and Yamamoto}}]{PhysRevB.85.075130}
\bibinfo{author}{\bibfnamefont{T.}~\bibnamefont{Byrnes}},
  \bibinfo{author}{\bibfnamefont{T.}~\bibnamefont{Horikiri}},
  \bibinfo{author}{\bibfnamefont{N.}~\bibnamefont{Ishida}},
  \bibinfo{author}{\bibfnamefont{M.}~\bibnamefont{Fraser}}, \bibnamefont{and}
  \bibinfo{author}{\bibfnamefont{Y.}~\bibnamefont{Yamamoto}},
  \bibinfo{journal}{Phys. Rev. B} \textbf{\bibinfo{volume}{85}},
  \bibinfo{pages}{075130} (\bibinfo{year}{2012}),
  \urlprefix\url{http://link.aps.org/doi/10.1103/PhysRevB.85.075130}.

\bibitem[{\citenamefont{Kasprzak}(2006)}]{Kasprzak:2006jt}
\bibinfo{author}{\bibfnamefont{J.}~\bibnamefont{Kasprzak}}, Ph.D. thesis,
  \bibinfo{school}{Universit{\'e} Joseph Fourier-Grenoble 1},
  \bibinfo{address}{Laboratoire de Spectrom{\'e}trie Physique-CNRS UMR 5588}
  (\bibinfo{year}{2006}).

\bibitem[{\citenamefont{Kasprzak
  et~al.}(2006{\natexlab{b}})\citenamefont{Kasprzak, Richard, Kundermann, Baas,
  Jeambrun, Keeling, Marchetti, Szyma{\'n}ska, Andr{\'e}, Staehli
  et~al.}}]{Kasprzak:2006mb}
\bibinfo{author}{\bibfnamefont{J.}~\bibnamefont{Kasprzak}},
  \bibinfo{author}{\bibfnamefont{M.}~\bibnamefont{Richard}},
  \bibinfo{author}{\bibfnamefont{S.}~\bibnamefont{Kundermann}},
  \bibinfo{author}{\bibfnamefont{A.}~\bibnamefont{Baas}},
  \bibinfo{author}{\bibfnamefont{P.}~\bibnamefont{Jeambrun}},
  \bibinfo{author}{\bibfnamefont{J.~M.~J.} \bibnamefont{Keeling}},
  \bibinfo{author}{\bibfnamefont{F.~M.} \bibnamefont{Marchetti}},
  \bibinfo{author}{\bibfnamefont{M.~H.} \bibnamefont{Szyma{\'n}ska}},
  \bibinfo{author}{\bibfnamefont{R.}~\bibnamefont{Andr{\'e}}},
  \bibinfo{author}{\bibfnamefont{J.~L.} \bibnamefont{Staehli}},
  \bibnamefont{et~al.}, \bibinfo{journal}{Nature}
  \textbf{\bibinfo{volume}{443}}, \bibinfo{pages}{409}
  (\bibinfo{year}{2006}{\natexlab{b}}).

\bibitem[{\citenamefont{Kasprzak
  et~al.}(2008{\natexlab{a}})\citenamefont{Kasprzak, Richard, Baas, Deveaud,
  Andr{\'e}, Poizat, and Dang}}]{Kasprzak:2008lh}
\bibinfo{author}{\bibfnamefont{J.}~\bibnamefont{Kasprzak}},
  \bibinfo{author}{\bibfnamefont{M.}~\bibnamefont{Richard}},
  \bibinfo{author}{\bibfnamefont{A.}~\bibnamefont{Baas}},
  \bibinfo{author}{\bibfnamefont{B.}~\bibnamefont{Deveaud}},
  \bibinfo{author}{\bibfnamefont{R.}~\bibnamefont{Andr{\'e}}},
  \bibinfo{author}{\bibfnamefont{J.-P.} \bibnamefont{Poizat}},
  \bibnamefont{and} \bibinfo{author}{\bibfnamefont{L.~S.} \bibnamefont{Dang}},
  \bibinfo{journal}{Phys. Rev. Lett.} \textbf{\bibinfo{volume}{100}},
  \bibinfo{pages}{067402} (\bibinfo{year}{2008}{\natexlab{a}}).

\bibitem[{\citenamefont{Kasprzak
  et~al.}(2008{\natexlab{b}})\citenamefont{Kasprzak, Solnyshkov, Andr{\'e},
  Dang, and Malpuech}}]{Kasprzak:2008mi}
\bibinfo{author}{\bibfnamefont{J.}~\bibnamefont{Kasprzak}},
  \bibinfo{author}{\bibfnamefont{D.~D.} \bibnamefont{Solnyshkov}},
  \bibinfo{author}{\bibfnamefont{R.}~\bibnamefont{Andr{\'e}}},
  \bibinfo{author}{\bibfnamefont{L.~S.} \bibnamefont{Dang}}, \bibnamefont{and}
  \bibinfo{author}{\bibfnamefont{G.}~\bibnamefont{Malpuech}},
  \bibinfo{journal}{Phys Rev Lett} \textbf{\bibinfo{volume}{101}},
  \bibinfo{pages}{146404} (\bibinfo{year}{2008}{\natexlab{b}}).

\bibitem[{\citenamefont{Richard et~al.}(2005)\citenamefont{Richard, Kasprzak,
  Romestain, Andr{\`e}, and Dang}}]{Richard:2005tw}
\bibinfo{author}{\bibfnamefont{M.}~\bibnamefont{Richard}},
  \bibinfo{author}{\bibfnamefont{J.}~\bibnamefont{Kasprzak}},
  \bibinfo{author}{\bibfnamefont{R.}~\bibnamefont{Romestain}},
  \bibinfo{author}{\bibfnamefont{R.}~\bibnamefont{Andr{\`e}}},
  \bibnamefont{and} \bibinfo{author}{\bibfnamefont{L.~S.} \bibnamefont{Dang}},
  \bibinfo{journal}{Phys Rev Lett} \textbf{\bibinfo{volume}{94}},
  \bibinfo{pages}{187401} (\bibinfo{year}{2005}).

\bibitem[{\citenamefont{Marchetti et~al.}(2006)\citenamefont{Marchetti,
  Keeling, Szyma\ifmmode~\acute{n}\else \'{n}\fi{}ska, and
  Littlewood}}]{PhysRevLett.96.066405}
\bibinfo{author}{\bibfnamefont{F.~M.} \bibnamefont{Marchetti}},
  \bibinfo{author}{\bibfnamefont{J.}~\bibnamefont{Keeling}},
  \bibinfo{author}{\bibfnamefont{M.~H.}
  \bibnamefont{Szyma\ifmmode~\acute{n}\else \'{n}\fi{}ska}}, \bibnamefont{and}
  \bibinfo{author}{\bibfnamefont{P.~B.} \bibnamefont{Littlewood}},
  \bibinfo{journal}{Phys. Rev. Lett.} \textbf{\bibinfo{volume}{96}},
  \bibinfo{pages}{066405} (\bibinfo{year}{2006}),
  \urlprefix\url{http://link.aps.org/doi/10.1103/PhysRevLett.96.066405}.

\end{thebibliography}
\end{document}